\documentclass{aa}
\usepackage{verbatim}
\usepackage{graphicx}
\usepackage{booktabs}
\usepackage{txfonts}
\usepackage{color}
\usepackage{ulem}
\usepackage{bm}
\usepackage{amsmath}

\usepackage[]{xcolor}

\usepackage{hyperref}
\hypersetup{
    colorlinks,
    linkcolor={blue},
    citecolor={blue},
    urlcolor={blue}
}

\begin{document}

   \title{The likelihood of not detecting cavity-carving companions in transition discs -- A statistical approach}
   \author{Enrico Ragusa\inst{\ref{unimi},\ref{unimiMat}}, Giuseppe Lodato\inst{\ref{unimi}}, Nicol\'as Cuello\inst{\ref{UGA}}, Miguel Vioque\inst{\ref{ESO}}, Carlo F. Manara\inst{\ref{ESO}},
   Claudia
   Toci\inst{\ref{ESO}}}

   \authorrunning{E. Ragusa et al. 2025}

   \institute{Dipartimento di Fisica, Università degli Studi di Milano, Via Celoria 16, 20133 Milano MI, Italy\label{unimi}
   \and
   Dipartimento di Matematica, Università degli Studi di Milano, Via Saldini 50, 20133, Milano, Italy \label{unimiMat}\\ \email{enrico.ragusa@unimi.it}
   \and
   Univ. Grenoble Alpes, CNRS, IPAG, 38000 Grenoble, France\label{UGA}
   \and
   European Southern Observatory, Karl-Schwarzschild-Strasse 2, 85748 Garching bei M\"unchen, Germany \label{ESO}}

   \date{Received XXX; accepted YYY}

  \abstract
{Protoplanetary discs with cavities, also known as transition discs, constitute $\sim$ 10\% of the protoplanetary discs at submillimmeter wavelengths. As one of several explanations, one hypothesis suggests that these cavities are carved by undetected stellar or planetary companions.  
 }
{We present a novel approach for quantifying the likelihood that a companion that carves the cavity in a transition disc is not detected because it is too close to the central star (small projected separation) or too faint to be resolved.}
{We generated two independent samples of stellar and planetary companions that were randomly oriented in the sky. We assumed a distribution of their eccentricity, mass ratio, and time-weighted orbital phases to study the statistical properties of the cavities they carve. We first calculated the likelihood that each companion in these samples appears at a certain projected separation $d$ relative to its semi-major axis $a_{\rm bin}$ ($d/a_{\rm bin}$). Then, we applied a disc truncation model to calculate the likelihood that each companion carves a cavity with a size $a_{\rm cav}$ relative to its semi-major axis $a_{\rm bin}$ and projected separation $d$, deriving distributions of $a_{\rm bin}/a_{\rm cav}$ and $d/a_{\rm cav}$.}
{We find that stellar companions carve cavities with sizes $a_{\rm cav}$ with a median about three times larger than their projected separation $d$ ($a_{\rm cav}\sim 3\, d$, and $a_{\rm cav}\sim 1.7\, d$ times for planets), but with a statistically significant tail ($\sim 50\%$) towards higher values ($a_{\rm cav}\gg 3\, d$). With this information, we estimated the likelihood that cavity-carving companions remain undetected because of projection effects when the system is observed with a spatial resolution
$\mathcal R$, $P(d<\mathcal R\,a_{\rm cav})$.}
{Using observational constraints on companion masses, we applied this framework to 13 well-known transition discs. We conclude that an undetected stellar companion is unlikely in 8 out of the 13 systems we considered, with 5 notable exceptions: AB~Aur, MWC~758, HD~135344B, CQ~Tau, and HD~169142. A planet, on the other hand, may have remained undetected in any of the transition discs we considered.}

   \keywords{Planet-disc interactions; Protoplanetary discs; (Stars:) binaries general; (Stars:) formation; (Stars:) pre-main sequence }

   \maketitle
%

\section{Introduction}\label{sec:introduction}

The significant advancements in our observational capabilities during the past decade revealed a large variety of features such as spirals, shadows, gaps, and other non-axisymmetric features in discs that surround forming stars (e.g. \citealp{long2018,andrews2018,cieza2021}, \citealp{bae2023} for a review).
Among them, roughly 10\% of the observed protoplanetary discs ($\sim 30\%$ of the brightest systems) present large ~10 –100 au dust and gas-depleted cavities that surround their forming stars (see \citealp{vandermarel2023} for a review).

Discs with cavities, often referred to as transition discs, belong to the brightest\footnote{Even though they also appear to be frequent around less luminous low-mass M-type stars \citep{shi2024}.} and most frequently studied sources in close star-forming regions \citep{pinilla2018,vandermarel2018,francis2020}. Their formation has been associated with stellar (e.g. \citealp{artymowicz1996,ragusa2017,price2018a}) or planetary (e.g. \citealp{ataiee2013,ovelar2013,regaly2017}) companions that interact with the disc. This interaction is a well-known mechanism for depleting the material in the co-orbital region of the companion. Other processes, such as grain growth, dead zones, and photoevaporation, have also been proposed for the formation of these cavities \citep[e.g.][]{dullemond2001,alexander2006,regaly2012,pinilla2016,ercolano2017,vandermarel2023,huang2024}.

Although the scenario involving companions appears to be widely invoked in this context, almost no confirmed detections are available in transition disc cavities, with the exception of PDS~70 \citep{keppler2018,muller2018} and HD142527, whose binary companion cannot have carved the observed cavity \citep{nowak2024}. This raises doubts about the companion scenario. Recently, \citet{vandermarel2021} placed observational upper limits on the companion masses in the cavities of several transition discs. The authors concluded that companions with mass ratios $q\gtrsim 0.05$ can be excluded in most systems at the location of the gas density minimum. However, it remains possible that companions might be hosted at smaller radii.

\begin{figure*}
   \centering   \includegraphics[width=\textwidth]{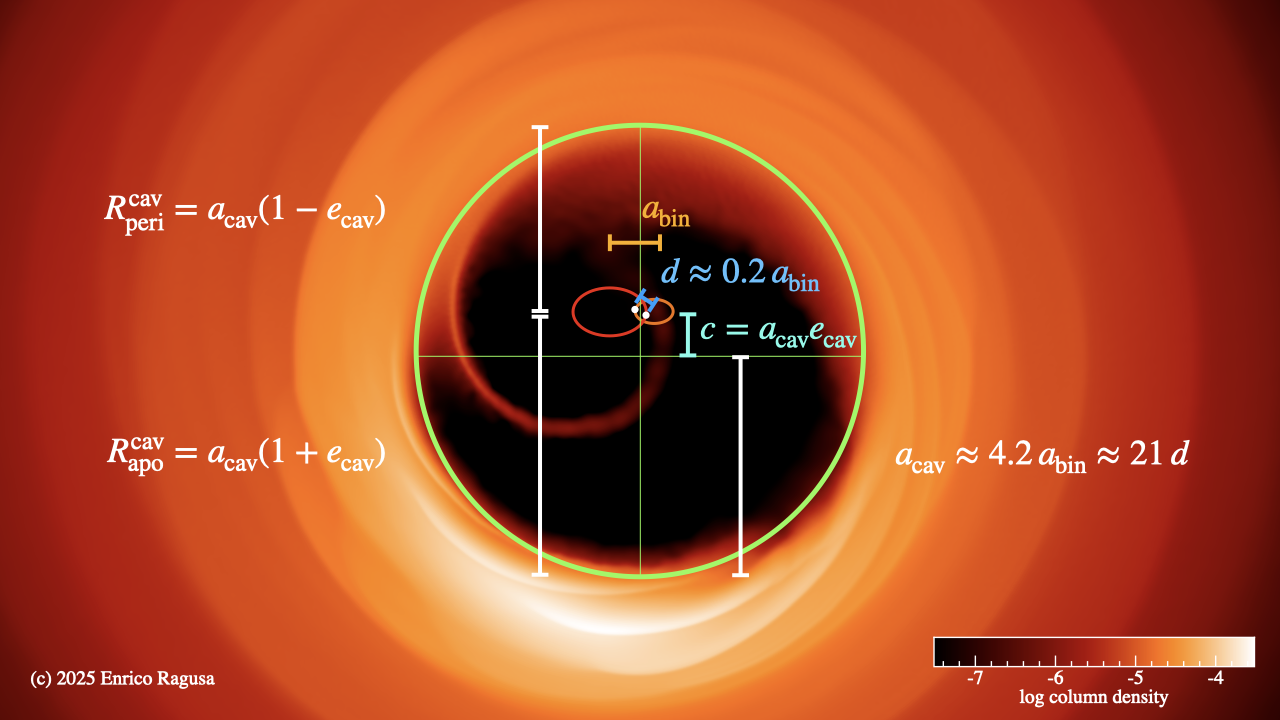}
   \caption{Snapshot of a numerical simulation of a circumbinary disc captured in the extreme situation in which the cavity size $a_{\rm cav}$ appears to be larger by a factor $\sim 21$ than the apparent separation $d$ of the binary. The snapshot was produced using the code \textsc{phantom} \citep{price2018a}. It shows a binary with a mass ratio $M_2/M_1=0.5$, semi-major axis $a_{\rm bin}=1$, and eccentricity $e_{\rm bin}=0.8$ close to its pericentre, that is, with a projected binary separation of  $d\sim 0.2 a_{\rm bin}$. The binary produces a cavity with an eccentricity $e_{\rm cav}\sim 0.2$ and a semi-major axis $a_{\rm cav}\sim  4.2\, a_{\rm bin}$ (consistent with the expected truncation radius by such a binary; see Sect. \ref{sec:truncprescr}), which is equivalent to $a_{\rm cav}\sim 21 \,d$. The two white dots mark the position of the two binary stars, and the orange and red ellipses show the orbits of the primary and secondary object, respectively. The binary spends far less time at its pericentre than at the apocentre. It is therefore unlikely that the binary is observed in this configuration, but it is still possible.}
\label{fig:simexample}%
\end{figure*}

Most theoretical works estimated a cavity size $a_{\rm cav}\sim 2\, \textrm{--\,}4 \,a_{\rm bin}$, where $a_{\rm bin}$ is the companion semi-major axis\footnote{Throughout the paper, we often refer to stellar and planetary companions as ``binaries''. We use the term in its broader definition of ``two gravitationally bound masses''.} (e.g. \citealp{artymowicz1994,pichardo2005,miranda2015,thun2017,zhang2018,hirsh2020,ragusa2020,dittman2024,penzlin2024}). Conclusions concerning the detectability of companions based on the observational detection limits mentioned above were typically drawn by comparing the semi-major axis of a putative companion $a_{\rm bin}$ of a certain mass with the detection limit at location $R=a_{\rm bin}$. This is equivalent to assuming an apparent separation of the binary of $d=a_{\rm bin}$. However, projection effects that arise because the binary is close to pericentre or to the inclination of the orbit in the plane of the sky can significantly reduce the projected (apparent) separation $d$ of the binary compared to its semi-major axis $a_{\rm bin}$. This implies that in a number of instances, $a_{\rm cav}\gg 4\,d$ and that in order to exclude the presence of companions in a cavity, the comparison with the detection limits also needs to account for the uncertain apparent separation of the putative binary. Figure \ref{fig:simexample} shows an extreme example from a numerical simulation. It shows an eccentric binary  at the pericentre of its highly eccentric orbit ($e=0.8$) carving a cavity that features $a_{\rm cav}\sim 21 \,d$, despite $a_{\rm cav}\sim 4\, a_{\rm bin}$. This causes the binary to appear extremely compact compared to the cavity, and there is a high probability that it remains undetected because the observational resolution in the area is poor (e.g. covered by a coronagraph).

In light of these considerations, we present a novel statistical approach for quantifying the likelihood that a planetary or stellar companion, that is assumed to be carving the cavity, remains undetected because it is located too close to the central star (small projected separation) or too faint to be resolved at the moment of observation. 
To do this, we produced two samples of companions (one sample for planetary companions, and the other sample for stellar companions) with arbitrarily distributed orbital properties. We studied the resulting projected separations $d$ relative to the binary semi-major axis $a_{\rm bin}$ ($d/a_{\rm bin}$, Sect. \ref{sec:binary sample}). Using a disc truncation prescription (Sect. \ref{sec:truncation}), we then studied the size of the cavity $a_{\rm cav}$ that each companion in the samples would carve in a disc. We calculated the distributions of the projected separations relative to the cavity sizes ($d/a_{\rm cav}$, Sect. \ref{sec:probabdistr}) and their cumulative distributions Sect. (\ref{sec:doacav1}). This quantity acts as the likelihood that a putative companion that carved the cavity remained undetected because the resolution is too low. We then used the observational constraints on the upper mass limits for the companion from \citet{vandermarel2021} to estimate the likelihood that the cavity of a set of real transition discs was carved by undetected stellar or planetary companions because the required sensitivity was lacking (Sect. \ref{sec:obs}). We discuss the results in Sect. \ref{sec:discussion} and the caveats of our analysis in Sect. \ref{sec:caveats}, and we draw our conclusions in Sect. \ref{sec:concl}.

\section{Samples and projected separations}\label{sec:binary sample}

The projected separation  $d$ for a binary with semi-major axis $a_{\rm bin}$, inclination $i$ in the plane of the sky, eccentricity $e$, longitude of pericentre $\varpi$, and true anomaly $f$ reads \citep{vanalbada1968}
\begin{equation}
    \frac{d}{a_{\rm bin}}=\frac{1-e^2}{1+e\cos(f)}\left[1-\sin(f+\varpi)^2\sin(i)^2\right]^{1/2}.\label{eq:doa}
\end{equation}

By assuming the distributions of  ($f,e,i,\varpi$) that are characteristic of a binary population, we can therefore generate the resulting distribution of $d/a_{\rm bin}$ for the population.

To create the planetary and stellar binary populations, we generated two samples of $N=8.5\times 10^5$ companions with orbital properties ($f,e,i,\varpi$) specified as follows. The two populations shared the same geometric assumptions for the orientation in space of the orbits: a uniform distribution in 3D space of the orbital inclinations $i$ and the pericentre longitude $\varpi$, which implies a uniform distribution of $\cos(i)$, that is, $\mathcal  P_i=\sin (i)/2$, with $i$ spanning $0<i<{\rm \pi}$, and a uniform distribution $\mathcal P_\varpi=1/2{\rm \pi}$ of the longitude of the pericentres between $0<\varpi<2{\rm \pi}$.

For both populations, the distribution of the true anomalies $\mathcal P_f$ needs to account for the fact that companions spend more time at the apocentre of their orbit than at the pericentre. This distribution for $\mathcal P_f$ was obtained by first noting that the binary angular velocity is ${\rm d}f/{\rm d}t=\Omega$. As a consequence, the fraction of time $dt$ that a binary spends in one orbit between a true anomaly $f$ and $f+{\rm d}f$ is
\begin{equation}
    \frac{{\rm d}t}{t_{\rm orb}}= \frac{\Omega_0}{\Omega}\frac{{\rm d}f}{2{\rm \pi}},
\end{equation}
where $t_{\rm orb}$ is the orbital time $t_{\rm orb}=2{\rm \pi}/\Omega_0$, and $\Omega_0=\sqrt{GM_{\rm bin}/a_{\rm bin}^3}$, where $M_{\rm bin}$ is the total mass of the binary. For an eccentric binary, $\Omega$ is
\begin{equation}
\Omega=\Omega_0\frac{[1+e\cos(f)]^2}{[1-e^2]^{3/2}}.
\end{equation}
This implies that $\mathcal P_f$ reads
\begin{equation}
   \mathcal P_f=\frac{\Omega_0}{2{\rm \pi}\Omega}= \frac{(1-e^2)^{3/2}}{2{\rm \pi}[1+e\cos(f)]^2}.
\end{equation}

The two populations differ in the distributions we used to generate the eccentricity $e$ and the binary mass ratio $q=M_2/M_1$, where $M_1$ and $M_2$ are the primary and secondary masses of the binary.

For the stellar binary population, we assumed that the distribution $\mathcal P_{e_{\rm bin}}$ of the eccentricity $e$ is uniform between $0\leq e<1$, while the distribution $\mathcal P_{q_{\rm bin}}$ of $q$ is uniform, with $q$ varying between $0.01\leq q\leq 1$. This qualitatively reproduces the distributions observed for solar-type field binary stars with separations ranging from few au to $\sim 100$ au \citep{duchene2013,moe2017,offner2023}. The choice of the lower limit $q=0.01$ for the stellar binary distribution is based on the commonly used $13M_{\rm J}$ lower limit for deuterium burning. This sets the threshold mass at which planets are typically considered to be brown dwarfs. For a solar-mass primary star, this limit translates into $q\gtrsim 0.01$.

\begin{figure}
   \centering
   \includegraphics[width=\columnwidth]{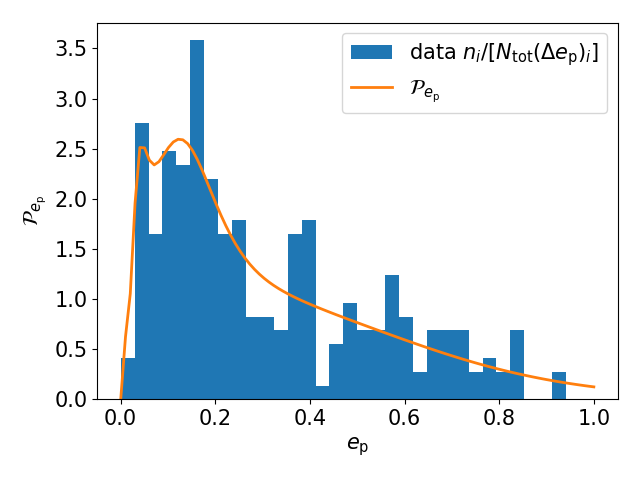}  \includegraphics[width=\columnwidth]{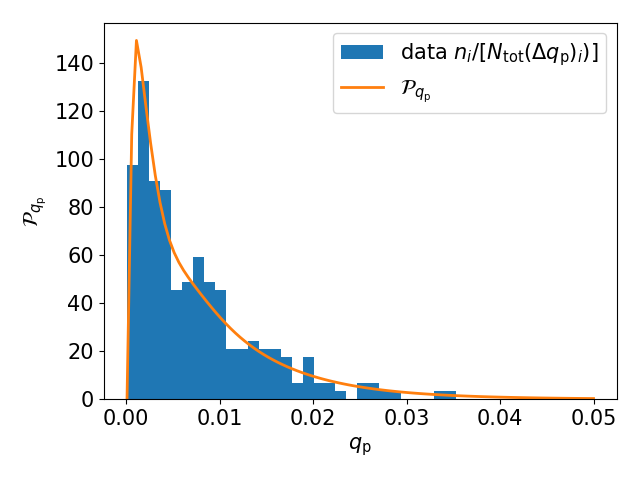}
   \caption{Probability distributions $\mathcal P_{e_{\rm p}}$ (top panel) and $\mathcal P_{q_{\rm p}}$ (bottom panel) for the eccentricity $e$ and the mass ratio $q$ of the planet population, respectively.}
\label{fig:planeteccq}%
\end{figure}

For the planet population, we used the distributions of the eccentricities $e$, $\mathcal P_{e_{\rm p}}$, and the mass ratios $q$, $\mathcal P_{q_{\rm p}}$, from the exoplanet population. We obtained them
through a kernel density evaluation (KDE) from the exoplanet dataset (\href{https://exoplanetarchive.ipac.caltech.edu}{NASA Exoplanet Archive}) with $a>2.5$ au and $q>5\times 10^{-4}$ . The lower limit on the mass ratio here constitutes a conservative estimate of the minimum mass required for gaps to open in the gas and dust density distributions \citep{dipierro2017,zhang2018}. The probabilities were obtained by performing the KDE on the $\log$ of $e$ and $q$ of the exoplanets and by exploiting the relations $\mathcal P_e=\mathcal P _{\log e}/e$ and $\mathcal P_q=\mathcal P _{\log q}/q$. A comparison between the exoplanet data and the distributions is shown in Fig. \ref{fig:planeteccq}. The eccentricity distribution appears to be a Rayleigh distribution, which is consistent with \citet{zhou2007}.
We note that we did not correct for observational bias. However, we do not expect observational bias to revert the general properties of $\mathcal P_{e_{\rm p}}$ and $\mathcal P_{q_{\rm p}}$. A few caveats that derive from our assumptions for the protoplanet population are discussed in Sect. \ref{sec:caveats}.

\begin{figure*}
   \centering
   \includegraphics[width=\columnwidth]{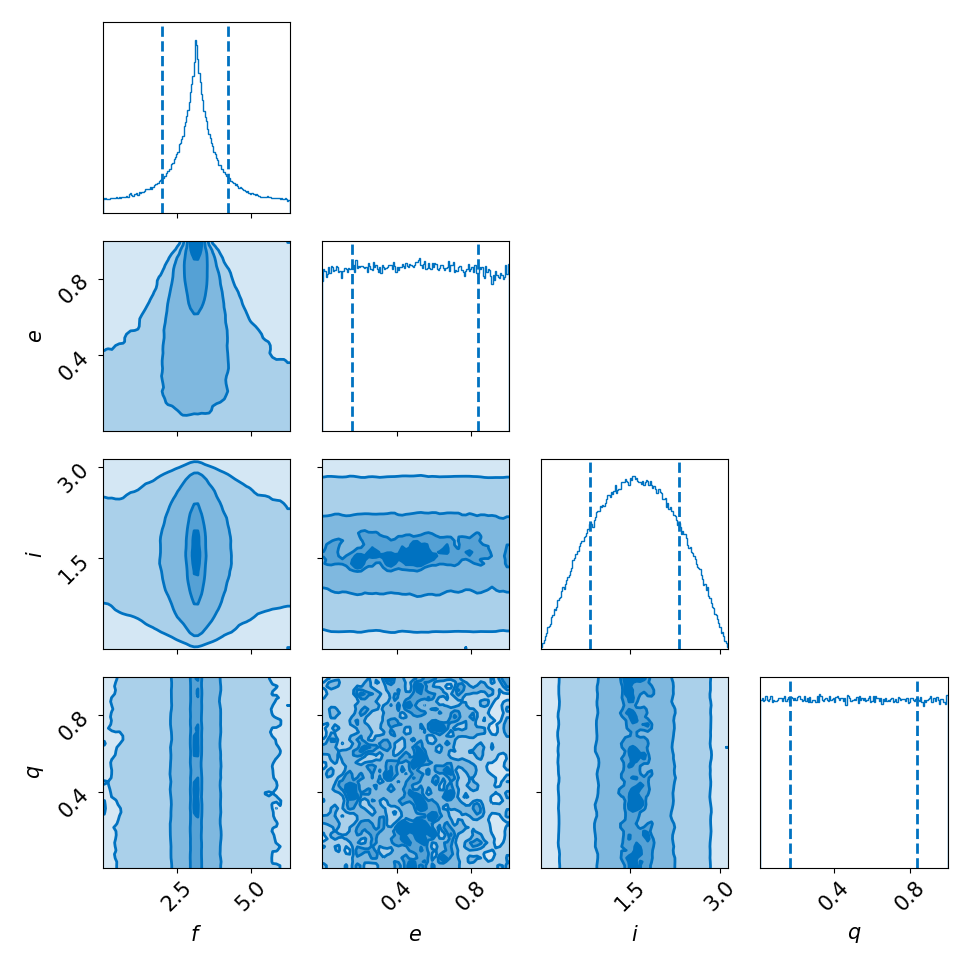}
   \includegraphics[width=\columnwidth]{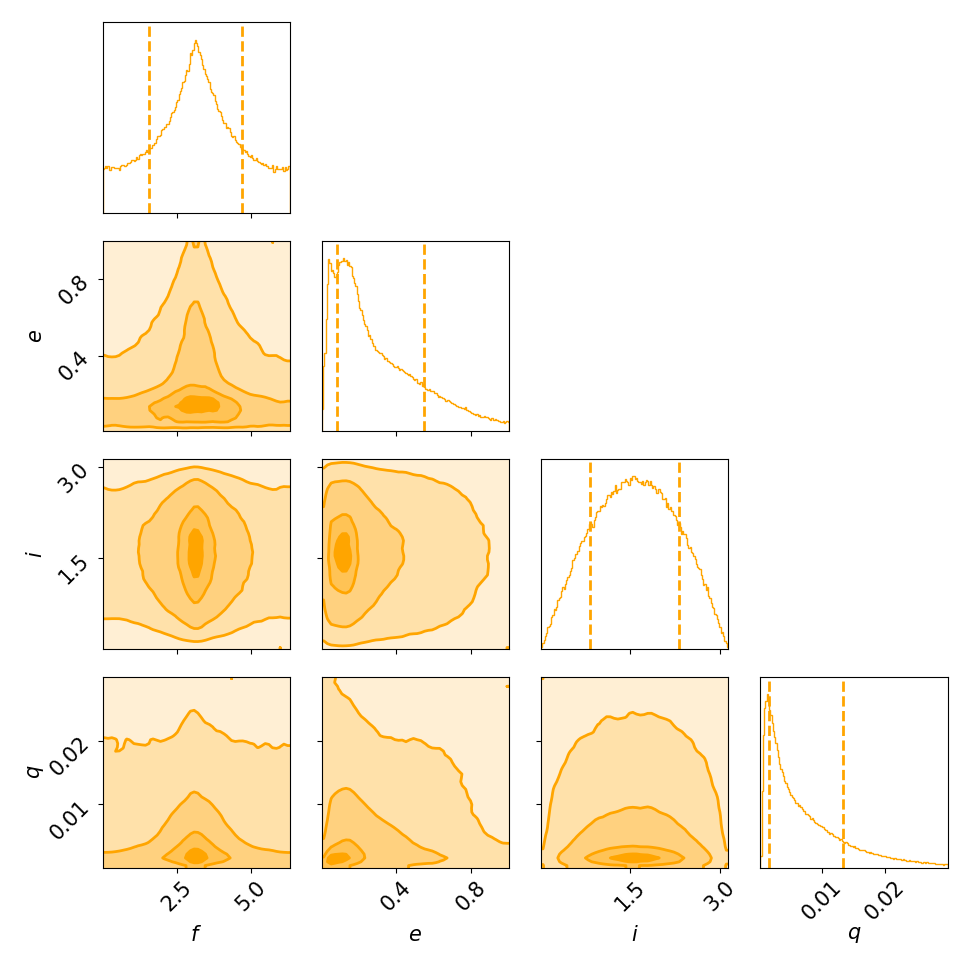}
   \caption{Statistical properties of the samples. Left panel: Corner plot distribution of the binary sample we generated for the variables $f$, $e$, $i$, and $q$ for the binary population. Right panel: Same as in the left panel, but for the planet population. We deliberately omitted $\mathcal P_\varpi$ because for the binary and planet populations, the longitude of the pericentre $\varpi$ is sampled from a uniform distribution between $0<\varpi< 2{\rm \pi}$. The vertical dashed lines represent intervals containing $16\%$ and $84\%$ of the sample ($2\sigma$).}
\label{fig:corner}%
\end{figure*}

\begin{figure}
   \centering
   \includegraphics[width=\columnwidth]{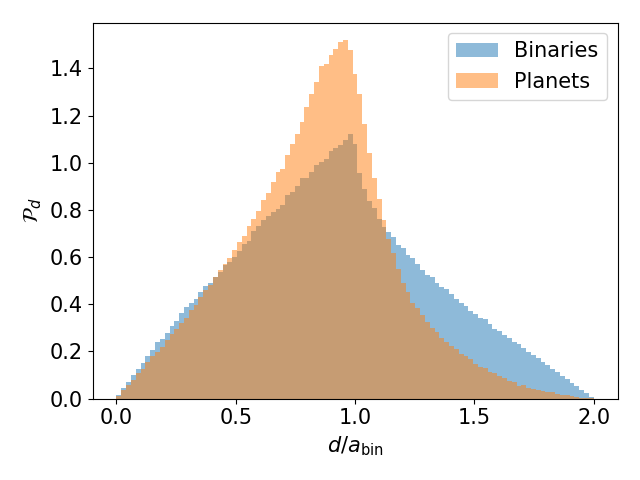}
   \caption{Resulting probability density function (pdf) distributions $\mathcal P_d$ of the quantity $d/a_{\rm bin}$ (Eq. \ref{eq:doa}), that is, the ratio of the projected separation of the binary $d$ and its semi-major axis $a_{\rm bin}$, for the stellar binary  (shaded blue pdf) and planetary population (shaded orange pdf). The probability density value was obtained by binning the values of $d/a_{\rm bin}$ and normalising each bin by $n_i/N_{\rm tot}\Delta x_i$, where $n_i$ is the number of counts in the bin, $N_{\rm tot}$ is the total number of elements in the sample, and $\Delta x$ is the bin size.}
\label{fig:histbin}%
\end{figure}

In Fig. \ref{fig:corner} we show the corner plot distributions of ($f,e,i,q$) of the companions in the two populations. The distribution of $\varpi$ is not plotted because it is uniform. In Fig. \ref{fig:histbin} we show the resulting $\mathcal P_{d/a}$ distribution of the quantity $d/a_{\rm bin}$ for both samples. We note that while $e$, $i$, $\varpi$, and $q$ are not correlated, there is a correlation between $e$ and $f$ , as evident in Fig. \ref{fig:corner}. This is a consequence of the dependence of $\mathcal P_f$ on $e$ and $f$ because the higher the binary eccentricity, the longer the time it spends at its apocentre.

The planetary and stellar binary distributions $\mathcal P_{d/a}$ both peak at a projected separation $d/a_{\rm bin}\sim 1$. However, large tails of projected separations $d$ smaller and larger than $a_{\rm bin}$ can be noted, with a slightly higher probability for $d<a_{\rm bin}$ in the planet and stellar binary populations. This effect is stronger in the planetary population: while the eccentricity contributes to producing both $d/a_{\rm bin}\lessgtr 1$, the inclination only reduces the observed projected separation when $i\neq 0$. This favours $d<a_{\rm bin}$ configurations in the planetary population, that is, configurations that are characterised by lower eccentricities. In Appendix \ref{appendix1} we compare the results obtained with our approach and those obtained by \citet{vanalbada1968}. The comparison confirms the perfect consistency between the two.

\section{Disc truncation}\label{sec:truncation}

In this section, we introduce a prescription for the disc truncation, based on which we calculated the cavity size $a_{\rm cav}$ carved by each companion in the samples presented in Sect. \ref{sec:binary sample} if it were surrounded by a protoplanetary disc.

To keep the discussion concise, we briefly introduce the process of disc truncation in Sect. \ref{sec:generalconsiderations}, and we then introduce in Sect. \ref{sec:truncprescr} the truncation prescription that is used throughout. Appendix \ref{sec:trunc2} provides an overview of the literature contextualising our choice for the truncation prescription, and Appendix \ref{sec:resononreso} contains additional information about the differences and similarities of the two types of truncation mechanisms (resonant and non-resonant).

\subsection{General considerations}\label{sec:generalconsiderations}

When a secondary companion mass (a planet or another star) is present in a protoplanetary system, its gravitational potential has two main effects on the surrounding material: (i) It excites waves at resonant locations in the disc (i.e. regions in which the orbital frequency of the companion and the disc have an integer commensurability) that deposit angular momentum and energy in it. This pushes the material (dust/gas) away from the co-orbital region (e.g. \citealp{goldreich1980,goodman2001}). (ii) It destabilises the orbits of the material and creates regions that are devoid of material (e.g. \citealp{rudak1981,pichardo2005}). When these mechanisms become sufficiently effective, the gravitational interaction leads to a change in the density structure of the disc.

When the mass ratio $q=M_2/M_1$ of the companion mass $M_2$ to the central star $M_1$ exceeds a certain threshold, a gap opens in the co-orbital region of the companion (for typical protoplanetary disc parameters $q>10^{-3}$ for a gas-gap opening \citealp{crida2006}, and  $q>10^{-4}$ for a dust-gap opening \citealp{dipierro2017}). The typical companions in this gap-forming mass regime are planets that can form characteristic gap features. Inside the gap, the planet is surrounded by a circumplanetary disc or envelope, while a stream of material connects the outer disc with the inner disc across the gap through the L1 and L2 Lagrange points.

For higher mass ratios ($q>0.04$), the companion starts to produce a cavity instead of a gap. The structure and dynamics of the disc become more complex and have three distinct components: two circumstellar discs, each of which surrounds the primary and secondary stars and is externally truncated by mutual gravitational interactions; and one circumbinary disc that encompasses both stars and is separated by a material-depleted region known as the cavity. The origin of this transition between the gap and cavity regimes is the lack of horseshoe and/or tadpole-type orbits, that is, stable orbits around the Lagrange points L4 and L5 \citep{murray1999} for mass ratios $q\gtrsim 0.04$.

Despite the substantial dynamical difference between gaps and cavities, numerical simulations of protoplanetary discs have shown that the inner edge of gaps can spread inward \citep{zhu2012,lambrechts2014,rosotti2016,ubeira2019} when the tidal barrier that is produced by the companion causes dust filtering (i.e. the largest dust grains cannot cross the companion orbit; e.g. \citealp{rice2006}) or reduce the gas-accretion rate across the gap, which prevents the inner disc from being refilled with fresh material. This can also produce cavity-like features in the planetary mass regime. Alternatively, a large cavity might be the result of the combined effects of multiple planets such as in the system PDS~70, where two massive planets carve the prominent cavity in the system \citep{keppler2018,bae2019,toci2020}. In these instances, the outermost planet regulates the distance from the cavity edge. In summary, planets and stellar companions produce cavities in discs. This process is also commonly referred to as disc truncation.

The mass and orbital properties of the companions determine the size of the cavity they carve, measured by its semi-major axis $a_{\rm cav}$. This is roughly equivalent to the radial extent of the cavity up to moderate values of the cavity eccentricity\footnote{The ratio of the semi-minor and semi-major axes is $b_{\rm cav}/a_{\rm cav}=\sqrt{1-e_{\rm cav}^2}$, so that for $e_{\rm cav}\lesssim 0.3$, we obtain $b_{\rm cav}\gtrsim 0.95\,a_{\rm cav}\sim R_{\rm cav}$.} $e_{\rm cav}$. Most commonly, $a_{\rm cav}$ is defined through the gas/dust density as the radial location at which the density $\Sigma$ becomes a fraction $\delta=10\%\,\textrm{--}\,50\%$ of the maximum density value $\Sigma_{\rm max}$ at the edge of the cavity, such that $\Sigma(a_{\rm cav})=\delta\,\Sigma_{\rm max}$ (e.g. \citealp{crida2006,duffell2015,kanagawa2018}). This is also known as the truncation radius.
The value of $a_{\rm cav}$ can be predicted analytically based on binary-disc interaction theory or numerically with hydrodynamic simulations.

\subsection{Truncation prescription in this work}
\label{sec:truncprescr}

We split our prescription into two separate mass regimes: the stellar companion regime ($q>0.01$), and the planetary regime ($q<0.01$). The prescriptions provide a smooth transition in the value of $a_{\rm cav}$ at $q=0.01$. We refer to Appendix \ref{sec:trunc2} for a thorough discussion of truncation mechanisms and their dependence on the system parameters.

For stellar mass companions, we used a truncation prescription based on the 3D stability of the three-body problem by \citet{georgakarakos2024}, which is part of the non-resonant family.
The empirical formula of these authors provides the critical semi-major axis of a test particle that orbits a binary with a mass ratio $0.01\leq q\leq1$. Below this critical value, the orbit is found to be unstable for all initial true anomalies. We assumed that this innermost orbit marks the gas-truncation radius. A comparison between this truncation prescription and the results for disc truncation by \citet{pichardo2005} can be found in Appendix \ref{appendix:comparison}. The general agreement with the numerical results is good.

The empirical formula from \citet{georgakarakos2024} expressed as a function of our relevant variables reads
\begin{align}
 \log_{10}\left(\frac{a_{\rm cav}}{a_{\rm bin}}\right)&=0.30889-0.26446\,M_{\rm lb}+ 0.09362\,i_{\rm d}+ \nonumber\\
 &+0.37426\,e_{\rm bin}+ 0.31306\,e_{\rm cav}-0.27007\,M^2_{\rm lb}+\nonumber\\
&-0.06102\,i^2_{\rm d}-0.09262\,e^2_{\rm bin}+0.19436\,M_{\rm lb}e_{\rm bin}+ \nonumber\\
&-0.18911\,i_{\rm d}e_{\rm bin}-0.05466\,M^3_{\rm lb}+0.06746\,M^2_{\rm lb}e_{\rm bin}+\nonumber\\
&+0.08715\,i^2_{\rm d}e_{\rm bin}+1.19488\,e^3_{\rm cav}, \label{eq:acav}
\end{align}
where $M_{\rm lb}=\log_{10}(q/q+1)$, with $q$  the binary mass ratio, the binary eccentricity $e_{\rm bin}$, the test particle inclination (in our case, the mutual binary-disc inclination $i_{\rm d}$), and the test particle eccentricity (in our case, the cavity eccentricity $e_{\rm cav}$). We refer to Appendix \ref{appendix:comparison} for an additional discussion of the advantages and motivations for the choice of Eq. (\ref{eq:acav}) as our truncation prescription.

\begin{figure}
   \centering
\includegraphics[width=\columnwidth]{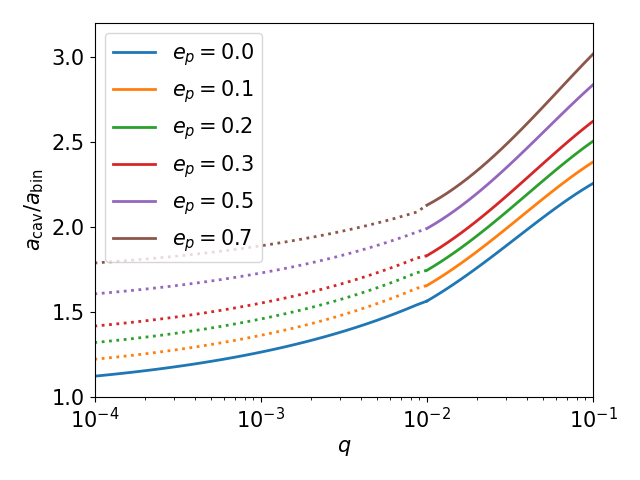}
   \caption{Prescription for $a_{\rm cav}$ that combines Eq. (\ref{eq:acav}) for $q\geq 0.01$ and (\ref{eq:acav_plan}) for $q<0.01$. The transition between the two prescriptions at $q=0.01$ is smooth. The different colours represent different values of the planet/binary eccentricity. The dotted lines for $q<0.01$ are plotted to highlight the transition between the two prescriptions.}
\label{fig:faprescr_plan}%
\end{figure}

For mass ratios $q<0.01$, that is, in the planetary mass regime, we adopted the following prescription:
\begin{equation}
    \frac{a_{\rm cav}}{a_{\rm bin}}=1+e_{\rm bin}+k \frac{R_{\rm Hill}}{a_{\rm bin}}\sqrt{1-e_{\rm bin}^2},\label{eq:acav_plan}
\end{equation}
where $R_{\rm Hill}=a_{\rm bin}\left(q/3\right)^{1/3}$ is the Hill radius, and $k=3.79$ is a constant that produces a smooth transition between the planetary and stellar truncation regimes across $q=0.01$, as shown in Fig.~\ref{fig:faprescr_plan}.

This prescription combines multiple features of dependence of the gap width on the planet properties. (i) The known scaling relation of gap widths with the Hill radius $\Delta=kR_{\rm Hill}$ (e.g. \citealp{rosotti2016}, as discussed in Appendix \ref{sec:trunc2}), to which it reduces when $e_{\rm bin}=0$ ($a_{\rm cav}=a_{\rm bin}+\Delta$). (ii) The scaling relation found by \citet{chen2021} that $a_{\rm cav}\propto 1+e_{\rm bin}$ when $a_{\rm bin}e_{\rm bin}\gtrsim R_{\rm Hill}$ (i.e. when the planet epicycle exceeds its Hill radius), which agrees qualitatively with orbital stability studies \citep{petrovich2015}. (iii) The factor $\sqrt{1-e_{\rm bin}^2}$ enables the smooth connection with the low-mass ratio regime of the \citet{georgakarakos2024} prescription (Fig. \ref{fig:faprescr_plan}). It can be interpreted as a factor that rescales the Hill radius to the average distance $R_p$ between the planet and star along the orbit, $a_{\rm p}\sqrt{1-e^2}=({\rm 2 \pi})^{-1}\int_0^{2\rm \pi}R_p{\rm d}f$ , or as the geometric average between the distance of apocentre and pericentre.
(iv) Finally, with proper tuning of $k$, Eq. (\ref{eq:acav}) produces a smooth transition with the \citet{georgakarakos2024} prescription at $q=0.01$ for all values of $e$ (as shown in Fig. \ref{fig:faprescr_plan}). This should be interpreted as an indication that the cavity size indeed depends on $q$ and $e_{\rm bin}$ also in the $q<0.01$ regime: the transition of one curve can be tuned, but the same value of $k$ does not necessarily produce a smooth transition for the other curves as well.

The prescription in Eq. (\ref{eq:acav_plan}) was developed under the assumption that the disc and planet are coplanar, and it does not depend on the mutual disc-planet inclination $i_{\rm d}$.  This is a reasonable assumption because most exoplanets show low mutual inclinations, which implies that they formed within the disc and continued to orbit close to the disc orbital plane. Similarly to Eq. (\ref{eq:acav}), we assumed that the prescription in Eq. (\ref{eq:acav_plan}) defines the truncation radius of the gaseous disc: by construction, $k=3.79$ ensures a smooth transition between the stellar (relevant for the gas) and planetary mass regime. In Sect. \ref{sec:obs} we discuss an empirical relation that links the size of the gas cavity with that of the dust cavity for a direct comparison with dust continuum sub-millimeter observations.

\section{Projected separation of the binary for a given cavity size. The distribution of $d/a_{\rm cav}$}\label{sec:probabdistr}

\begin{figure}
   \centering
   \includegraphics[width=\columnwidth]{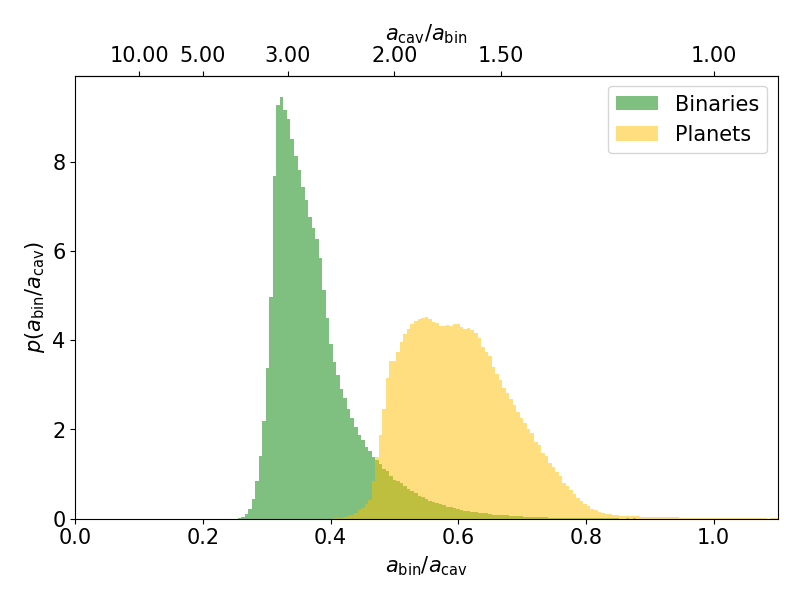}\\
   \includegraphics[width=\columnwidth]{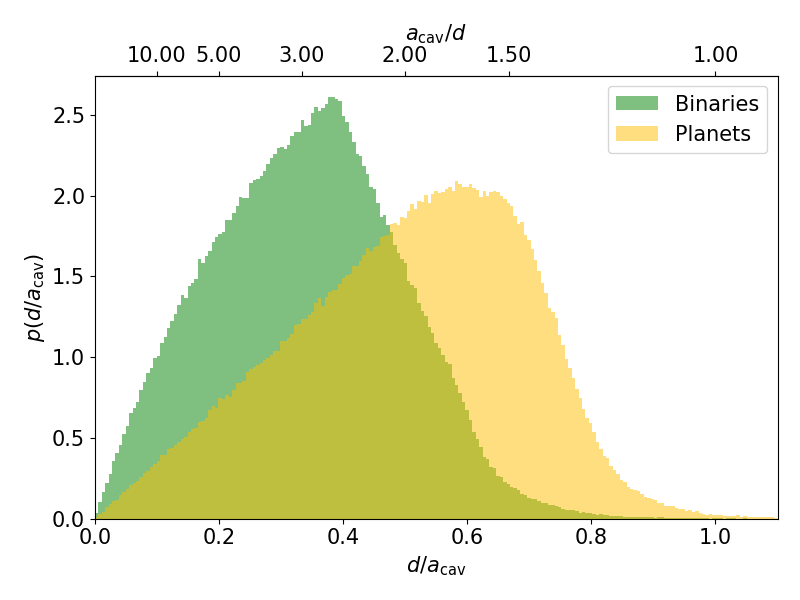}
   \caption{Probability density distribution of $a_{\rm bin}/a_{\rm cav}$ (top panel) and $d/a_{\rm cav}$ (bottom panel). The probability density was obtained by binning the values of $d/a_{\rm cav}$ and normalising each bin by $n_i/N_{\rm tot}\Delta x_i$, where $n_i$ is the number of counts in the bin, $N_{\rm tot}$ is the total number of elements in the sample, and $\Delta x$ is the bin size.}
\label{fig:histcav}%
\end{figure}

For each element in the samples that we generated in Sect. \ref{sec:binary sample}, we calculated the values of $a_{\rm cav}/a_{\rm bin}$ by applying Eq. (\ref{eq:acav}) or (\ref{eq:acav_plan}), depending on their mass ratio.
To do this, we assumed the mutual binary/planet-disc inclination $i_{\rm d}$, which does not coincide with the inclination in the plane of the sky of the companions $i$. For the stellar binary case, we prescribed that $i_{\rm d}$ was a shuffled version of the inclination $i$ array, meaning that binaries and discs have random mutual inclinations (we also tested $i_{\rm d}=0$, i.e. coplanar discs, but there was no significant difference).  For the planetary case, as explained in Sect. \ref{sec:truncprescr}, Eq. (\ref{eq:acav_plan}) does not depend on $i_{\rm d}$ by definition. This implies that the planet and disc are assumed to be coplanar, $i_{\rm d}=0$. As mentioned previously, we considered this to be a reasonable assumption for the planetary case.
The top panel of Fig. \ref{fig:histcav} shows the resulting distribution of $a_{\rm bin}/a_{\rm cav}$, from which we note that stellar companions carve cavities with sizes of $a_{\rm cav}\sim 2\,\textrm{--}\, 4\,a_{\rm bin}$, and planets carve cavities with sizes of $a_{\rm cav}\sim 1.5\,\textrm{--}\, 2\,a_{\rm bin}$.

We calculated the ratio of the projected separation $d/a_{\rm bin}$ obtained in Sect. \ref{sec:binary sample} and $a_{\rm cav}/a_{\rm bin}$ to obtain the values of $d/a_{\rm cav}$ for our samples. Fig. \ref{fig:histcav} shows the resulting distribution of $p(d/a_{\rm cav})$, which provides information about the frequency with which companions with projected separation $d$ carve cavities with a size of $a_{\rm cav}$ (see Fig. \ref{fig:histcav}).

The planetary and stellar $d/a_{\rm cav}$ distributions we obtained highlight that cavities are most likely to have projected separations $a_{\rm cav}\sim 3\,d $ (i.e. $ d/a_{\rm cav}\sim 0.33$) for stellar companions and $a_{\rm cav}\sim 1.7\, d$ (i.e. $d/a_{\rm cav}\sim 0.7$) for planets, which agrees with the respective $a_{\rm bin}/a_{\rm cav}$ distributions. However, long tails that extend to $d/a_{\rm cav}\sim 0$ (i.e. $a_{\rm cav}\gg 3\, d$) can be observed at smaller projected separations. This highlights the statistical importance of companions at small projected separations.

The physical origin of the difference between the two populations lies in the smaller regions of unstable orbits that surround planets compared to stellar binaries. As a result, the last stable orbits around planets are closer to the co-orbital region than those of stellar binaries. This means that planets carve out smaller cavities than stellar binaries on average, with the same semi-major axis. From another perspective, planets exert a lower tidal torque on the disc than stellar binaries.

In the following sections, we use the statistical information on the projected separations relative to the cavity sizes to study the likelihood that an undetected companion remains undetected within the cavity of transition discs because the spatial resolution is too low (Sect. \ref{sec:doacav1}), under the assumption that they cause the formation of the cavity. We then extend the analysis by considering limitations of the observational sensitivity (Sect. \ref{sec:obs}).

\subsection{The likelihood that cavity-carving companions remain undetected because the projected separation is small}\label{sec:doacav1}

We started by exploring the case of direct-imaging observations and ignored the possibility that companions might be detected with other techniques, such as radial velocity variations or interferometry (e.g. sparse aperture masking or VLTI-GRAVITY).

A companion that carves the cavity in a transition disc might remain undetected when the projected separation of the binary is too small to be resolved, because the spatial resolution of the instrument is too low, or when it is obscured by a coronagraph during the observations.

We calculated the cumulative distribution of $p(d/a_{\rm cav})$ to determine the likelihood of this scenario,
\begin{equation}
    P\left(d/a_{\rm cav}<\mathcal{R}\right)=\int^{\mathcal{R}}_{0}p(x)dx.\label{eq:cumuld}
\end{equation}
The cumulative probability $P(d/a_{\rm cav}<\mathcal R)$ quantifies the percentage of elements in our samples that have $d/a_{\rm cav}<\mathcal R$. This quantity informs us about the fraction of companions in the sample that would not be resolved because they are obstructed by a coronagraph with a size of $\mathcal R$ ($\mathcal R a_{\rm cav}$ is the coronagraph size in physical units), or more generally, because the resolution over a spatial length $\mathcal R$ is too low, and that might be hidden within the cavity.

\begin{figure}
\centering
\includegraphics[width=\columnwidth]{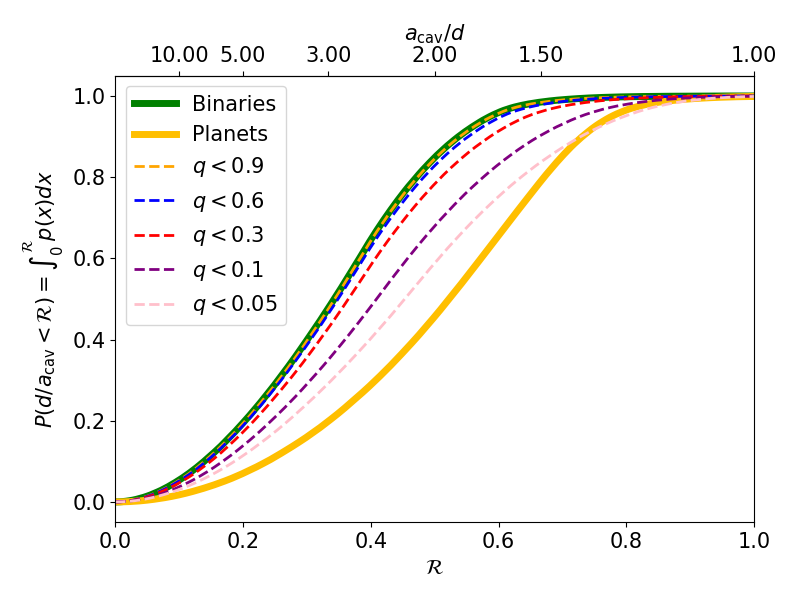}
   \caption{Cumulative distributions $P(d/a_{\rm cav}<\mathcal{R})$ calculated using Eq. (\ref{eq:cumuld}). The solid green and yellow lines show the cumulative distributions for the stellar binary and planetary populations. The dashed lines show distributions obtained by applying a cut $q<q_{\rm max}$ in the stellar binary regime and renormalising the total distribution to $P(d/a_{\rm cav}<1)=1$ to show the impact of $q$ on the cumulative distribution.}
\label{fig:cumuldist}%
\end{figure}

In Fig. \ref{fig:cumuldist} we show these cumulative distributions for the planet and binary samples. We also studied the cumulative distributions after imposing a maximum  mass ratio $q_{\rm max}$ and applying cuts in the binary sample, that is, we removed companions with $q>q_{\rm max}$ from the sample (see Sect. \ref{sec:obs} for realistic $q_{\rm max}$ from real observations).
We renormalised the distributions after the cuts to have $P(+\infty)=1$. This whole procedure is equivalent to changing the initial distribution from which the mass ratios are sampled, and it highlights that different assumptions on the upper limit $q_{\rm max}$ of the mass ratio affect the likelihood.
For a fixed value of $\mathcal R$, a lower $q_{\rm max}$ reduces the probability $P$ for $d/a_{\rm cav}<\mathcal R$. This effect becomes progressively stronger for $q_{\rm max}\lesssim 0.1$ because $a_{\rm cav}$ only mildly depends on the binary mass ratio for $q>0.1$ (we refer to Appendix \ref{sec:trunc2} for a discussion of the dependence of $a_{\rm cav}$ on the system parameters).

We also note that although the shape of the distributions for $q$ and $e$ for generating the planet and stellar companion samples differ strongly, the cumulative distribution for the planet population appears to be a natural extension of the stellar companion with $q_{\rm max}<0.05$, except for some differences in the tails. This suggests that the assumptions on the distribution of $e$ made to generate the samples play a marginal role in determining $P(d/a_{\rm cav})$ as long as moderate eccentricities are allowed in the sample, while $q_{\rm max}$ appears to be the key parameter.

\subsection{Comparison with the transition disc population}\label{sec:obs}

In this section, we extend the previous discussion to include sensitivity detection limits and provide an example of how the statistical approach to disc truncation we presented can be applied to observations. In particular, under the hypothesis that the cavities of transition discs are carved by companions, we calculate the likelihood that they remain undetected, considering the observational detection limits. We note that this is different from the probability that a binary or a planet is present within the cavity, as this analysis does not give us any information about alternative formation mechanisms for cavities. A low likelihood allows us to exclude our hypothesis that a hidden companion carves the cavity, but a high likelihood does not mean that a companion is expected to be found in the cavity.

\subsubsection{Sample and definition of the observational truncation radius}
We used the sample of transition discs from \citet{vandermarel2021}, who collected companion mass detection limits from the literature from sparse aperture masking, coronaph studies, and from lunar occultation. We select 13 systems with cavities from their sample, that is, we excluded systems with dust continuum emission from the central region (no sub-millimeter cavity) and PDS~70, which we know to host two protoplanets that carve its cavity \citep{bae2019,toci2020}.
We did not exclude HD142527, although its companion with $d\sim 13$ au means that it is a circumbinary disc, because recent constraints on its orbit ($a_{\rm bin}\sim 10$ au) appear to suggest that it cannot carve the large cavity of the system \citep{nowak2024}. This implies that a third unseen body might be present in the system. For such a third body, the constraints from the known binary orbital parameters ($q\sim0.1$, $e_{\rm bin}=0.47$, $a_{\rm bin}=10.8$, $i_{\rm d}=68^\circ$, \citealp{nowak2024}) suggest that it has a semi-major axis $\gtrsim 30$ au in order to have a stable orbit.

We first defined the observational truncation radius $a_{\rm cav}^{\rm obs}$ through a direct observable. For this purpose, we chose the radius of the dust ring $R_{\rm mm}$ (referred to as $R_{\rm dust}$ by \citealp{vandermarel2021}), which is defined as the location of the maximum dust emission, and for which a relation of the type $a_{\rm cav}^{\rm obs}=\epsilon R_{\rm mm}$ is reasonably expected to obtain the size of the gas cavity -- this is required so that we can apply our truncation prescription.

We found that $\epsilon=0.75$ produces a value of $a_{\rm cav}^{\rm obs}(R_{\rm mm})$ that compares well with the theoretical value $a_{\rm cav}$ from Eq. (\ref{eq:acav}) and (\ref{eq:acav_plan}) using as input parameters the real orbital parameters of companions in the four systems in which the orbits of companion were constrained: PDS~70 \citep{wang2021}, V892~Tau \citep{long2021}, IRAS~04158+2805 \citep{ragusa2021}, and GG~Tau \citep{keppler2020,toci2024} (see Tab. \ref{tab:finetuning} for the details of the comparison).

To support our empirical choice of the relation $a_{\rm cav}^{\rm obs}(R_{\rm mm})$ and for the fine-tuning of $\epsilon=0.75$, we also note the following:
\begin{enumerate}[(i)]
\item Assuming that the dust ring traces the gas pressure maximum, the characteristic radius used to define the edge of the cavity is usually the location at which the density reaches a fraction of its maximum value ($0.1\,\textrm{--}\,0.5\times \Sigma_{\rm max}$; e.g. \citealp{crida2006}). The characteristic length scale for the density gradient of a stable gap cannot be shorter than the disc vertical scale height, which implies that the edge of the gas cavity must be at a distance of at least $\gtrsim H$ from the dust density peak. By definition, this constrains $1-\epsilon> H/R$. \\

\item The scaling of $a_{\rm cav}^{\rm obs}(R_{\rm mm})$ agrees qualitatively with the relation between radius of the gas component and $R_{\rm CO}$ and $R_{\rm mm}$ discussed in \citet{facchini2018}, where $R_{\rm CO}/R_{\rm mm}=\epsilon \sim 0.6\,\textrm{--}\, 0.85$. \\

\item Using $\epsilon> 0.75$, that is, making the prescription more constrained, implies larger $a_{\rm cav}^{\rm obs}$ gas cavities. For these to be carved, companions with larger $a_{\rm bin}$ are required. This would make companions more easily detectable and would lower the likelihood that a companion remains undetected. As a consequence, the likelihood of hidden companions in a system that is already unlikely to host them does not increase simply because 
$\epsilon$ has been adjusted to a higher value. As a result, our choice of $\epsilon$ is conservative.
\end{enumerate}

\begin{table*}
\center
\caption{Comparison between observational $a_{\rm cav}^{\rm obs}$ and theoretical $a_{\rm cav}$ for a few systems with known orbital properties of the companions.}
\begin{tabular}{l|cccccccc}
\hline\hline
Name & $R_{\rm mm}$ [au] & $a_{\rm cav}^{\rm obs}=0.75\times R_{\rm mm}$ [au] & $e_{\rm bin}$ & $a_{\rm bin}$ [au] & $q$ & $i_{\rm d}$ [$^\circ$] & $a_{\rm cav}$ [au] \\
\hline
\\[-9pt]
PDS~70$^1$& 74 & 55 & 0.04 & 34 & 0.007 & 0. & 51\\
V892 Tau$^2$& 27 & 20 & 0.27 & 7.1 & $\sim 1$ & 8 & 20 \\
IRAS 04158+2805$^3$ & $240$ & 180 & $\sim 0.7$ & $\sim 55$  & $\sim 1$ & $\lesssim 20$ & 185 \\
GG Tau$^{4,5}$ & 220 & 165 & $0.2\,\textrm{--}\, 0.4$ & $50\,\textrm{--}\, 60$ & 0.77 &  $10^\circ \,\textrm{--}\, 30^\circ$ & 160\\
\hline
\end{tabular}
\tablefoot{ We defined $a_{\rm cav}^{\rm obs}=\epsilon R_{\rm mm}$, where $\epsilon=0.75$, $R_{\rm mm}$ is the observed radius of the centre of the dust ring at the edge of the cavity, and the theoretical $a_{\rm cav}$ was obtained from Eq. (\ref{eq:acav}) and (\ref{eq:acav_plan}) using the orbital parameters constrained from observations. [$^1$]\citealp{wang2021} (the reported parameters refer to the dynamically stable solutions of the outer planet, PDS~70c). [$^2$]\citealp{long2021}. [$^3$]\citealp{ragusa2021}. [$^4$]\citealp{keppler2020}. [$^5$]\citealp{toci2024}.}\label{tab:finetuning}
\end{table*}

\subsubsection{Companion mass detection limits and likelihood}\label{sec:obslikel}

The companion mass detection limits collected by \citet{vandermarel2021} are still currently valid, to our knowledge. The authors collected detection limits in the literature that were mainly obtained based on two methods\footnote{\label{footnote:IRS48}For IRS48 \citep{vandermarel2021} reports a detection limit obtained through lunar occultation from \citet{simon1995}; we report it among the sparse aperture upper limits that are relevant for $R<0.2$ arcsec. However, we note that this source is known to be deeply embedded in the cloud, and the conversion from flux sensitivity to mass upper limits might be unreliable \citep{calcino2019}.}: (i) coronagraph studies, which provided upper detection limits $q_{\rm cor}$ on the companion mass ratio outside the coronagraph area at radii $R_{\rm  cor}\sim 0.1\textrm{--}\,0.2$ arcsec; (ii) when available, sparse aperture masking, which provides an upper detection limit $q_{\rm sp}$ on the companion mass ratio at radii of typically $R_{\rm sp}\lesssim 0.2$ arcsec. Therefore, we introduced $q_{\rm max}(R)$ detection limits as continuous functions of the distance from the centre of the system $R$. This combines the upper limits that were obtained with different methods in each system.

We defined $q_{\rm max}(R)$ based on a visual estimation from the continuous functions reported in Fig. 4 of \citet{vandermarel2021}. For each detection curve in \citet{vandermarel2021}, we defined two piecewise detection curves: one $q_{\rm max}^{+}$ constitutes an optimistic detection limit, and the other $q_{\rm max}^{-}$ constitutes a conservative detection limit.
The optimistic $q_{\rm max}^{+}$ reads
\begin{equation}
    q_{\rm max}^{+}(R) =\min \left\{
\begin{aligned}
    q_{\rm sp}^+, &&   0\leq R\leq R_{\rm sp}\\
    q_{\rm cor}^+, &&  R_{\rm cor}\leq R\leq a_{\rm cav}^{\rm obs}
\end{aligned}\right\},\label{eq:qradii1}
\end{equation}
where $q_{\rm sp}^+$ and $q_{\rm cor}^+$ are the maximum values of the detection curves in the sparse masking ($0\leq R\leq R_{\rm sp}$) and coronagraph regions ($R_{\rm cor}\leq x\leq R_{\rm cav}$), respectively; $R_{\rm sp}$ is the outer radius of the sparse masking, and $R_{\rm cor}$ is the coronagraph radius. The $\min(...)$ function is required because in some cases $R_{\rm sp}>R_{\rm cor}$.
The conservative $q_{\rm max}^-$ reads
\begin{equation}
    q_{\rm max}^{-}(R) =\min \left\{
\begin{aligned}
    q_{\rm sp}^-, &&   0\leq R\leq R_{\rm sp}\\
    q_{\rm cor}^-, &&  R_{\rm cor}\leq R\leq a_{\rm cav}^{\rm obs}
\end{aligned}\right\},\label{eq:qradii2}
\end{equation}
where $q_{\rm sp}^-$ and $q_{\rm cor}^-$ is the minimum-mass value of the detection curves in the sparse masking and coronagraph regions, respectively.
The values of $q_{\rm sp}^+$, $q_{\rm sp}^-$, $q_{\rm cor}^+$, $q_{\rm cor}^-$, $R_{\rm sp}$, and $R_{\rm cor}$ are reported in Tab. \ref{tab:transition discs}.

Under the hypothesis that transition disc cavities are carved by companions, we obtained the likelihood that a companion remains undetected. We took the distributions obtained in Sect. \ref{sec:binary sample} and removed from the sample the parameter configurations that for each observed system would already be ruled out based on the detection limits prescribed by Eq. (\ref{eq:qradii1}) and (\ref{eq:qradii2}), using the parameters in Tab. \ref{tab:transition discs}. Specifically, for each system we considered, we removed elements from the planet and binary samples if $q>q^\pm_{\rm max}(R)$ using $R=d$.

We calculated the cumulative distribution\footnote{To avoid confusion, we remark that the difference between $\mathcal R$ and $R$ is that the first is a non-dimensional parameter that compares with $d/a_{\rm cav}$, while $R=\mathcal Ra_{\rm cav}$ is in physical units and compares with $d$.} $P(d<R)$ using Eq. (\ref{eq:cumuld}) from these restricted samples. We plot the results in Fig. \ref{fig:trdiscs}. Each system is described by two $P(d<R)$ curves that refer to $q^+_{\rm max}$ (solid) and $q_{\rm max}^-$ (dotted), connected by a colour-shaded area. In this instance, $P(d<R)$ represents the likelihood that companions with a projected separation $d<R$ remain undetected in the disc cavity, under the assumption that companions are responsible for its formation. The shaded area thus defines a confidence interval for the likelihood between the upper limits of the optimistic ($q_{\rm max}^+$) and conservative ($q_{\rm max}^-$) mass ratio.

For each system, we defined separations in arcseconds and $a_{\rm cav}=a_{\rm cav}^{\rm obs}$ for our samples. To maintain consistency with the original exoplanet distribution, we adjusted the mass ratios of the planetary sample to $q/M_\star$, where $M_\star$ is the mass of the primary star in the system (the original $q$ assumes a primary star of $1\, {\rm M}_\odot$).

Additionally, for systems HD~142527, IRS~48, and MWC~758, we calculated the distribution of $d/a_{\rm cav}$ using the values $e_{\rm cav}=(0.3;0.3;0.1)$, respectively \citep{dong2018,kuo2022,garg2022,yang2023}.

System HD~142527 hosts a known binary that is too compact to carve the cavity \citep{nowak2024}. This implies that a tertiary companion in the system might be carving the cavity. We applied an additional cut to exclude companions with $d<30$ au. This represents the innermost stable circular coplanar orbit\footnote{Obtained using Eq. (\ref{eq:acav}) with $q=0.1$, $e_{\rm bin}=0.47$, $a_{\rm bin}=10.9$ au \citep{nowak2024}, and assuming a circular coplanar orbit of the tertiary companion.} around the binary. We accounted for this effect by including $q_{\rm sp}^\pm=0$ and $R_{\rm sp}=30$ for this system (see Tab. \ref{tab:transition discs}).

Finally, we defined optimistic and conservative values of $P(d<R=a_{\rm cav})$ for binaries and planets as $\mathcal L^\pm_{\rm bin}$ and $\mathcal L^\pm_{\rm pl}$, respectively, and we report them in the last two columns of Tab. \ref{tab:transition discs}. These quantities represent the total fraction of companions in our samples that satisfy the resolution and sensitivity criteria for remaining undetected within the whole cavity area. Therefore, they indicate the total likelihood of not detecting companions that are responsible for the formation of the cavity.

\begin{table*}[]
\center
\caption{Properties and detection limits of the transition discs considered in this work.}
\begin{tabular}{l|cccccccccccc}
\hline\hline
Name      &$M_\star$ [${\rm M}_\odot$] & $a_{\rm cav}^{\rm obs}$ [au] & $q_{\rm sp}^+$ & $q_{\rm sp}^-$   & $R_{\rm sp}$ [au] & $q_{\rm cor}^+$ & $q_{\rm cor}^-$  & $R_{\rm cor}$ [au] & $D$ [pc] & $\mathcal L_{\rm bin}^{\pm}$ & $\mathcal L^{\pm}_{\rm pl}$\\
\hline
\\[-9pt]
IRS~48    &2  & 52   & 0.075 & 0.075  & 100   & 0.05 &0.025  & 22 & 121 & 0.06 -- 0.05 & 1 -- 1 \\
HD~142527$^{1}$  &1.7 & 135   & 0 & 0 & 30 & 0.023 & 0.006  & 13 & 140 & 0.01 -- 0 & 0.91 -- 0.69 \\
AB~Aur$^{2}$    &2.6  & 127   & - &- &-   & -   & -    & -  &  144 & 1 -- 1 & 1 -- 1 \\
MWC~758   &1.7  & 37   & - & - &-  & 0.02 &0.01  & 15 & 200 & 0.72 -- 0.71 & 1 -- 0.94\\
HD~135344B$^{3}$ &1.4 & 38   & - & - &-  & 0.07 &0.01  & 15  & 140  & 0.64 -- 0.61 & 1 -- 0.9\\
SR~21    &2.1   & 27    & 0.18 &0.04  & 30& 0.01 &0.005 & 20 & 140 & 0.17 -- 0.03 & 1 -- 0.98\\
CQ~Tau    &1.7  & 37   & - & - & -  & 0.023 &0.01   & 25  & 100 & 1 -- 0.98 & 1 -- 0.98 \\
DoAr~44   &1.4  & 35   & 0.3 &0.03  & 30   & 0.07 &0.06 & 15  & 146 & 0.3 -- 0.02 & 1 -- 1\\
J1604-2130 &1 & 63   & 0.04 &0.015  & 20   & 0.01 &0.002 & 15  & 150 & 0 -- 0 & 0.77 -- 0.28\\
LkCa~15   &1.3  & 56   & 0.023 &0.015  & 25    & 0.015 &0.007  & 15 & 140 & 0.01 -- 0 & 0.96 -- 0.79\\
Sz~91     &0.6  & 70   & - & - &-  & 0.08 &0.015 & 15  & 159 & 0.27 -- 0.21 & 1 -- 0.73\\
HD169142  &1.7 & 19   & - & - &-   & 0.02 &0.01  & 12 & 117 & 1 -- 1 & 1 -- 1\\
DM~Tau    &0.5  & 19   & 0.06 &0.015  & 20    & - & - &- & 140 & 0.05 -- 0 & 1 -- 0.63\\
\hline
\end{tabular}
\tablefoot{$M_\star$ is the mass of the primary star in the system; $a^{\rm obs}_{\rm cav}$ is the semi-major axis of the gas cavity, obtained using $a^{\rm obs}_{\rm cav}=0.75\times R_{\rm mm}$, with $R_{\rm mm}$ the location of the dust cavity ring, as reported by \citet{vandermarel2021}; $q_{\rm sp}^\pm$ and $q_{\rm cor}^\pm$ are conservative and optimistic companion mass-ratio detection limits (Eq. \ref{eq:qradii1} and \ref{eq:qradii2}) from sparse aperture masking and from coronagraph studies, respectively, based on the minimum and maximum values in the detection curves in \citet{vandermarel2021}; and $R_{\rm sp}$, and $R_{\rm cor}$ are sparse aperture masking and coronagraph reference radii (see Eq. \ref{eq:qradii1} and \ref{eq:qradii2}). $\mathcal L^\pm_{\rm bin}$ and $\mathcal L^\pm_{\rm pl}$ are conservative and optimistic cumulative probabilities at the cavity edge ($P[d<R=a_{\rm cav}]$) for binaries and planets, respectively. These quantities represent the total likelihood that a companion remains undetected within the cavity under the hypothesis a companion carves it. [$^{1}$] The detection limits for HD142527 are for a tertiary undetected companion that carves the cavity. The detected binary in the system is too compact to carve the $a_{\rm cav}=135$ au cavity \citep{nowak2024}; $q_{\rm sp}^\pm=0$ up to $R_{\rm sp}=30$ are placed to account for orbital stability considerations around the binary in the system. [$^{2}$] For AB Aur, detection limits on the Pa$\beta$ planet accretion tracer are available \citep{biddle2024,currie2024}, but were not converted into upper mass limits. [$^{3}$]  For HD135344B, the newer upper limits presented by \citet{stolker2024} appear to be less conservative (higher upper limits) at large radii than those adopted by \citet{vandermarel2021}. We retained the more conservative values from \citet{vandermarel2021}.}
\label{tab:transition discs}
\end{table*}

\section{Discussion}\label{sec:discussion}

\begin{figure*}
\centering
\includegraphics[width=\textwidth]{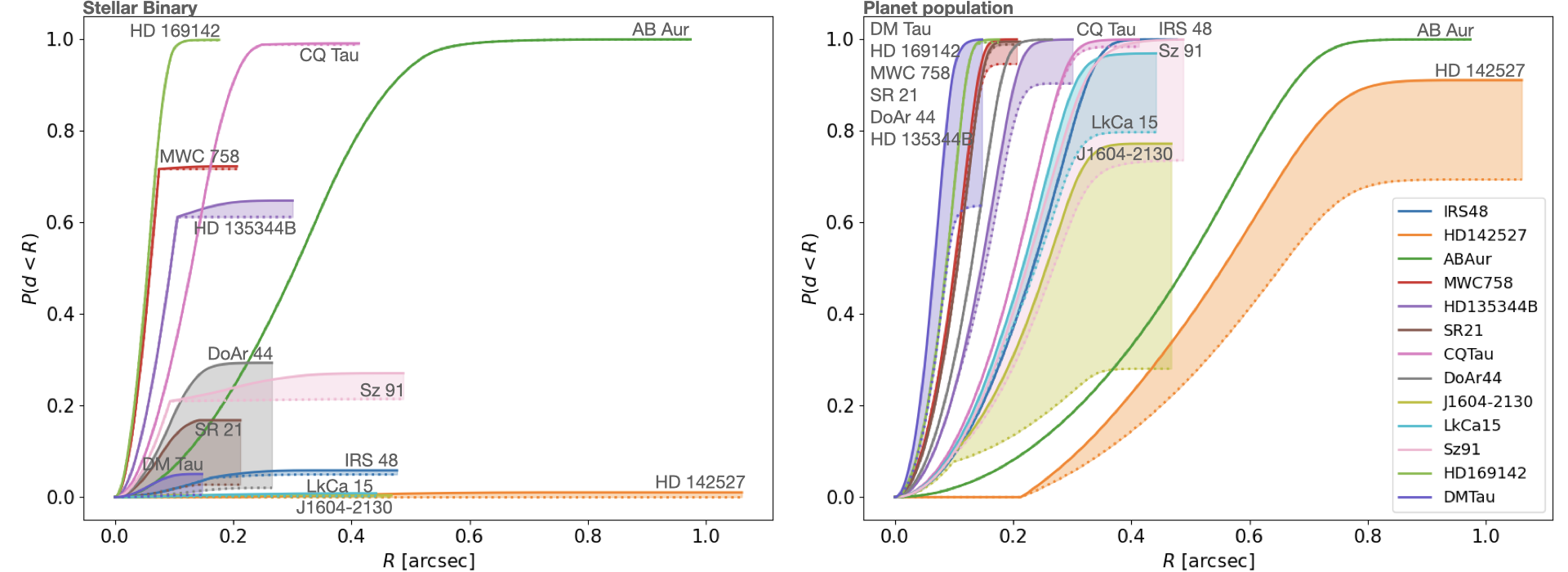}
   \caption{Likelihood $P(d< R)$ of non-detection of binary (left) or planetary mass (right) companions within the cavity of a set of observed transition discs as a function of radius under the hypothesis that companions caus the cavity to form (obtained as described in Sect. \ref{sec:obslikel}). For each system, the plot shows two curves that are joined by a colour-shaded area. The solid curve is associated with the optimistic ($q_{\rm max}^+$, Eq. \ref{eq:qradii1}) companion detection limits, and the dotted curve is associated with the conservative ($q_{\rm max}^-$, Eq. \ref{eq:qradii2}) companion detection limits. The colour-shaded area is a confidence interval between the two upper limits of the mass ratio. The end of each line marks the location of the cavity edge $R=a_{\rm cav}^{\rm obs}$.   Under the assumption that a companion forms the cavity, $P(d<R)$ represents the likelihood that a companion with a projected separation $d < R$ remains undetected in the disc cavity. For example, for the system CQ~Tau, assuming a companion is carving the observed cavity, $P(d<R=0.2)\sim 80\%$ represents the likelihood that it goes undetected if it has projected separation $d<0.2$ arcsec.
   The value of $P(d<R=a_{\rm cav}^{\rm obs})$, that is the value of $P$ at the end of each line, represents the total likelihood in the whole cavity area. We label the total likelihood for the conservative and optimistic upper limits of binaries and planets as $\mathcal L^\pm_{\rm bin}$ and $\mathcal L^\pm_{\rm pl}$, respectively, and report them in Tab. \ref{tab:transition discs}.
   Additional discussion can be found in Sect. \ref{sec:discussion}. As discussed in Sect. \ref{sec:obs}, we remark that a low likelihood allows us to exclude our hypothesis that a companion is responsible for the cavity; in contrast, a high likelihood does not mean that a companion is expected to be found in the cavity, but that under the hypothesis a companion is there carving the cavity high are the chances it goes undetected.
   }
\label{fig:trdiscs}%
\end{figure*}

Fig. \ref{fig:trdiscs} shows that the optimistic and conservative upper limits on the mass ratios at various distances from the central star mean that a stellar binary companion is unlikely in 8 of the 13 transition discs that we considered, with a likelihood $0\%<\mathcal L^\pm_{\rm bin}<30\%$. This low likelihood, even for the optimistic detection limits, leads us to conclude that undetected stellar binary companions that would carve the cavity can be safely excluded in these systems. However, for 5 systems -- AB~Aur (for which we could not find mass upper limits), MWC~758, HD~135344B, CQ~Tau, and HD~169142 -- the optimistic and conservative detection limits both show that a fraction $\mathcal L^\pm_{\rm bin}>60\%$ of the configurations of the stellar sample might carve the observed cavity even though they remain undetected. All these systems have been proposed to host planetary or stellar companions -- either putative or speculative -- within their cavities. (e.g. AB~Aur, \citealp{poblete2019,boccaletti2020,currie2022}; MWC~758, \citealp{dong2018}; HD~135344B, \citealp{stolker2016}, CQ~Tau, \citealp{ubeira2019}; and HD~169142, \citealp{fedele2017,toci2020b,poblete2022}).

In contrast, in all but one system, the majority of companions in the planetary sample would apparently remain undetected. The conservative and optimistic likelihoods both range between $70\%<\mathcal L_{\rm pl}^\pm<100\%$, including HD142527, for which, as mentioned above, a putative tertiary planetary companion has been suggested to carve the large cavity. System J1604-2130 instead has a likelihood $28\% <\mathcal L_{\rm pl}^\pm<77\%$, which makes it less likely that a planet that carves the observed cavity remains undetected, but it remains far from being ultimately ruled out

In general, we note that the nominal values of the detection limits are sufficiently low to exclude almost any stellar binary companion in regions that were explored through sparse aperture and coronagraph techniques. With the current detection limits, stellar companions are only possible in regions without applied limits because they were covered by the coronagraph and lacked sparse aperture masking limits. In contrast, the available sensitivity is typically only sufficient to detect the most massive planets with $M_{\rm p}\gtrsim 5\,\textrm{--}\,10 M_{\rm J}$, which constitute a small fraction of the planet population we assumed (the masses of $\sim 84\%$ of the planetary sample range between $5\times 10^{-4}\, \,{\rm M}_{\rm J}<M_{\rm p}\lesssim 13 \,{\rm M}_{\rm J}$). This leaves many possible planet configurations in the sample that can be hosted in transition discs, but cannot be detected.

\section{Caveats and assumptions}\label{sec:caveats}

The results we presented  depend on the choices and assumptions about (i) the planet and stellar binary populations, (ii) the observational upper limits on the companion masses (that might change with future observations), and (iii) the truncation prescriptions.

Concerning (i), more accurate estimates for the planetary and stellar populations can be obtained by refining the distributions we used to generate our samples. The stellar and planetary samples were both generated starting from considerations that are mostly applicable to evolved systems (field binaries and the exoplanet population). Furthermore, the planet distributions were obtained from the raw data in the NASA Exoplanet Archive without accounting for observational biases when extrapolating the properties of the planetary population. Finally, the mutual inclinations between discs and stellar binaries were taken to be randomly oriented for simplicity (uniform distribution of $\cos(i_{\rm d})$), while it might be more appropriate to assume a bimodal coplanar/polar distribution for stellar binaries (e.g. \citealp{aly2015,zanazzi2018, cuello2019b,martin2019}). However, the distributions we adopted constitute a reasonable approximation of the general properties of stellar and planetary populations. To support this statement, as noted in Sect. \ref{sec:doacav1}, the cumulative distribution of the stellar population in Fig. \ref{fig:cumuldist} appears to become progressively similar to the distribution for the planet population for decreasing values of $q_{\rm max}$, even though the underlying distributions for $e$ and $q$ are profoundly different. By experimenting with different distributions, we found that the key ingredients for shaping the qualitative behaviour of $P(d/a_{\rm cav}<\mathcal R)$
are (i) the value of $q_{\rm max}$, (ii) an at least moderate eccentricity for the elements in the sample ($e_{\rm bin}\gtrsim 0.15$), and (iii) the geometric projection in the plane of the sky. For these reasons, we therefore do not expect that a fine-tuning of the stellar and planetary populations will significantly change our conclusions (although it might be interesting).

Concerning (ii), as discussed in Sect. \ref{sec:obs}, our choices for mass-ratio detection limits for stellar and planetary companions constitute reasonable upper limits based on observations. Future observations are expected to further reduce the $P(d/a_{\rm cav}<\mathcal R)$ cumulative likelihood. However, some additional considerations in this regard deserve further discussion. Because the typical $q_{\rm max}\sim 0.005\,\textrm{--}\,0.01$, almost any stellar companion would be detected in spatial regions in which these limits apply ($R> 0.1$'')\footnote{The mass detection limits depend on the model of planet formation that is invoked. The limits reported here are consistent with the lowest end of masses that are detectable with SPHERE $M\gtrsim 5 \,{\rm M_{\rm J}}$. For a colder model, the lowest detectable mass would typically reach $M\gtrsim 15 \,{\rm M_{\rm J}}$ \citep{asensiotorres2021}.}. The stellar companions that survive have projected separations in regions where no limits have been placed because of the coverage of the coronagraph or insufficient resolution. In contrast, cuts with these limits only remove the most massive planetary companions. This issue has the following implications. On the one hand, we do not expect changes in the likelihood for stellar companions unless upper limits become available in currently unexplored regions.
On the other hand, the likelihood of planetary companions is not expected to change unless the current detection limits are lowered by a factor $5\,\textrm{--}\,10$, which implies a $q_{\rm max}\sim 10^{-3}$. This would allow the cuts to exclude a larger fraction of the planetary sample. The future advent of the Extremely Large Telescope and observations from the James Webb Space Telescope will enhance the resolution and sensitivity of planet detection campaigns through direct imaging. As a general word of caution, we note that foreground or disc extinction for systems deeply embedded in the cloud (e.g. IRS~48) may reduce the observed luminosity of potential companions, possibly making them not detectable even with the most sensitive instruments.

Concerning (iii), we consider our truncation prescription for the planetary and stellar companions to be reasonable (an additional discussion of how the prescription compares to other results for disc truncation can be found in Appendix \ref{appendix:comparison}). Fig. \ref{fig:histcav} shows that the distributions of $a_{\rm bin}/a_{\rm cav}$ indicate cavity sizes that typically lie in the range $2\, a_{\rm bin}\lesssim a_{\rm cav}\lesssim 4\, a_{\rm bin}$ for stellar binaries and $1.2\, a_{\rm bin}\lesssim a_{\rm cav}\lesssim 2\, a_{\rm bin}$ for planets. This is perfectly consistent with the extremal values that are typically expected for companions in the mass regime of our populations. In general, prescriptions that predict smaller cavities reduce the likelihood for stellar companions. This does not affect the planetary case strongly because planets would not be detectable in any case unless a significant reduction of $q_{\rm max}$ also occurs, as discussed above. Similar considerations also apply to the prescription for $a_{\rm cav}^{\rm obs}=R_{\rm mm}(1-\epsilon)$, for which a value of $\epsilon< 25\%$ would further reduce the likelihood. This would again be mostly effective on stellar companions and would be inclined to further reduce the likelihood.

We finally remark that even though $P$ is low, by construction, $P(d/a_{\rm cav}< \mathcal R)>0$ implies that some companions within the sample that survive after the cuts can in fact carve the observed cavities without being detected. In these cases, the presence of a companion should be considered unlikely, but not impossible. We also note that because binaries spend a very limited fraction of their orbital timescale close to their pericentre, the most compact configurations are relatively short-lived compared to the most extended ones. This was taken into account in the distribution we used to generate the true anomalies. The likelihood is expected to change, however, when more than one observation of the same system is performed after a time frame that permitted the companion to reach a less compact configuration.

\section{Conclusion}\label{sec:concl}

We presented a novel statistical approach for determining the likelihood that a binary or a planet remains undetected within the cavity of a transition disc under the hypothesis that a companion causes the cavity to form. In this approach, we combined upper limits of companion masses at different locations within the cavity and the probability that each companion is observed with a specific projected separation in the plane of the sky because of its orbital configurations and orientation in 3D space.

To do this, we created two samples, assuming reasonable distributions for the mass and orbital properties of stellar binaries and planets. These samples also accounted for the probability of orbital phases because companions spend more time at the orbit apocentre than at pericentre. We studied the resulting distributions of their projected separations $d$ relative to their semi-major axes $a_{\rm bin}$ ($d/a_{\rm bin}$) and to the cavity sizes $a_{\rm cav}$ that they are expected to carve if they were surrounded by a protoplanetary disc ($d/a_{\rm cav}$).
Our conclusions from this analysis are listed below.

(i) The planetary and stellar binary populations both have distributions of $d/a_{\rm bin}$ that peak at $\langle d/a_{\rm bin}\rangle\sim 1$. However, long tails for $d/a_{\rm bin}< 1$ and $d/a_{\rm bin}> 1$ can be observed in Fig. \ref{fig:histbin}. In general, this implies that some caution is required before the projected separation $d$ of a companion is used as a proxy value of its semi-major axis $a_{\rm bin}$.

(ii) A significant fraction of stellar binaries ($\sim 50\%$) produce cavities with $a_{\rm cav}>3\, d$ ($d/a_{\rm cav}<0.3$; see Fig. \ref{fig:cumuldist}), although for most systems, $a_{\rm cav}\sim 3\, a_{\rm bin}$ (see Fig. \ref{fig:histcav}). Similarly, planetary companions have a long tail with $a_{\rm cav}\sim 3\, d$ ($\sim 20\%$, for $d/a_{\rm cav}<0.3$) of configurations and a maximum likelihood for $a_{\rm cav}\sim 1.7\, d$ (i.e. $d/a_{\rm cav}\sim 0.6$).

(iii) Based on our statistical study that considered available upper detection limits (see Fig. \ref{fig:trdiscs}), we conclude that within the cavities of the systems we examined, undetected cavity-carving companions should be considered unlikely in 8 out of 13 systems, with a likelihood $\mathcal L^\pm_{\rm bin}<30\%$. This means that $\gtrsim 70\%$ of the companions in the stellar population sample in these 8 systems would be bright and separated enough from the central star to be observationally detected. These 8 systems are: IRS~48 (see footnote$^{\ref{footnote:IRS48}}$ for a word of caution about this system), HD~142527, SR~21, Sz~91, DoAr~44, DM~Tau, LkCa~15, and J1604+2130. However, 5 notable exceptions stand out with $\mathcal L_{\rm bin}^\pm>60\%$, namely: AB~Aur, MWC~758, CQ~Tau, HD~135344B, and HD~169142. In these systems, only a few of the possible configurations ($\lesssim 40\%$ in the worst case) can be ruled out based on the upper limits of companion detection.
In contrast, for the planetary sample, undetected planets remain potentially good candidates ($\mathcal L_{\rm pl}^\pm>80\%$) to carve the observed cavities of all transition discs except for one. The exception is J1604-2130, which has a likelihood of $28\% <\mathcal L_{\rm pl}^\pm<77\%$. However, this value of $\mathcal L_{\rm pl}^\pm$ implies a moderate likelihood that an undetected planet carved the cavity, although this possibility is far from being ultimately excluded. We recall that for HD~142527, the likelihood refers to the presence of a third companion in addition to the known binary, which has been shown to be too compact to have carved the observed cavity \citep{nowak2024}. As discussed in Sect. \ref{sec:obs}, we remark that the likelihood we refer to throughout the paper is the likelihood that companions remain undetected within the cavity of transition discs under the hypothesis that a companion carves it. We note that this is different from the probability that a binary or a planet is present within the cavity.

(iv) Since the most compact configurations are short-lived (since they are close to pericentre), the companion might reach a more extended detectable configuration after a relatively short time. Although the likelihood we discussed took the time spent by each binary in different orbital phases into account, it referred to one single observation. Therefore, it might be good to observe the system again after a sufficiently long period of time, which would allow the companion to reach more extended configurations. This would surely increase the detection probability if the sensitivity is adequate to observe it in its new location.
We finally remark that a low but not vanishing
$\mathcal L_{\rm bin}$ implies that the configurations for cavities are unlikely, but not impossible.

This work constitutes the first systematic statistical approach to evaluating the likelihood that putative companions that carve cavities in transition discs remain undetected.
At the time of observations, companions might be found in a configuration with a compact projected separation in the plane of the sky.
For this reason, we encourage the community to always rely on a statistical approach that accounts for this possibility before excluding the presence of companions from an observation. This will be particularly relevant with the new detection limits that are provided by the James Webb Space Telescope and the advent of the Extremely Large Telescope.

\begin{acknowledgements}

We thank the referee and the editor for their comments. ER thanks Nikolaos Georgakarakos for fruitful discussion about Eq. (\ref{eq:acav}) and orbital stability. ER acknowledges financial support from the European Union's Horizon Europe research and innovation programme under the Marie Sk\l{}odowska-Curie grant agreement No. 101102964 (ORBIT-D). ER also acknowledges the European Southern Observatory for hosting a three-month secondment within the ORBIT-D project, during which part of this project was developed. GL has received funding from the European Union's Horizon 2020 research and innovation program under the Marie Sklodowska-Curie grant agreement No. 823823 (DUSTBUSTERS) and from PRIN-MUR 20228JPA3A. NC acknowledges funding from the European Research Council (ERC) under the European Union Horizon Europe programme (grant agreement No. 101042275, project Stellar-MADE). CFM is funded by the European Union (ERC, WANDA, 101039452). Views and opinions expressed are however those of the author(s) only and do not necessarily reflect those of the European Union or the European Research Council Executive Agency. Neither the European Union nor the granting authority can be held responsible for them.

This research has made use of the NASA Exoplanet Archive (Fig. \ref{fig:planeteccq}), which is operated by the California Institute of Technology, under contract with the National Aeronautics and Space Administration under the Exoplanet Exploration Program.
Fig. \ref{fig:simexample} was created using \textsc{splash} \citep{price07a}. All the other figures were created using \textsc{matplotlib} python library \citep{hunter2007}.
\end{acknowledgements}

\bibliographystyle{aa}
\bibliography{biblio}

\appendix

\section{Comparison with Van Albada (1968)}\label{appendix1}

\begin{figure}
   \centering   \includegraphics[width=\columnwidth]{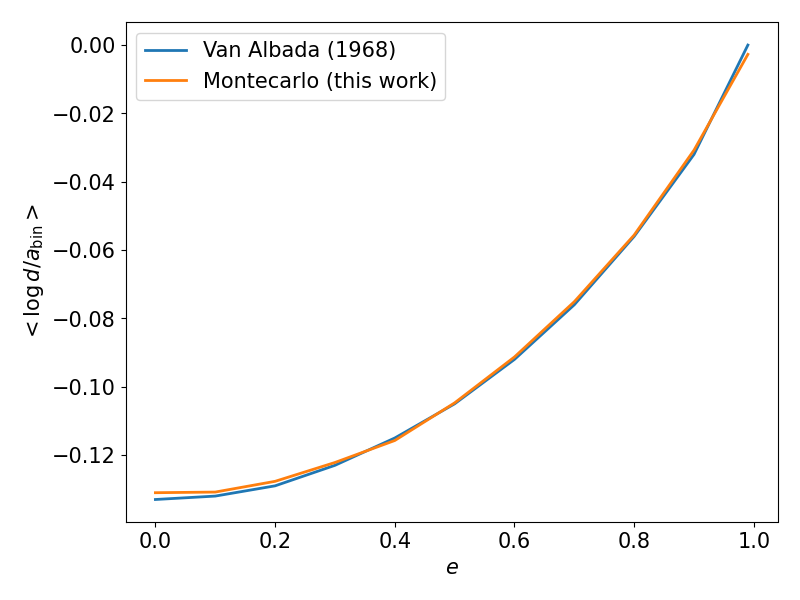}
   \caption{Comparison of $\langle\log(d/a)\rangle$ from \citet{vanalbada1968} (blue curve) and this work (orange). Note that $\langle \log (d/a_{\rm bin}) \rangle\neq \log \langle(d/a_{\rm bin} \rangle)$.}
\label{fig:vanalbada}%
\end{figure}

In Fig. \ref{fig:vanalbada}
we compare our Monte Carlo approach with the analytical calculations of \citet{vanalbada1968} for the quantity $\langle \log (d/a_{\rm bin})\rangle$ obtained using fixed values of eccentricity. This result is presented here to show that our Monte Carlo approach is equivalent to the purely analytical approach used by \citet{vanalbada1968}. However, our Monte Carlo approach enables more flexibility in the management of the distributions that generate the sample. Before any consideration about how Fig. \ref{fig:vanalbada} compares with the results discussed in Sect. \ref{sec:binary sample}, one should keep in mind that $\langle \log (d/a_{\rm bin}) \rangle\neq \log (\langle d/a_{\rm bin} \rangle)$.

\section{Overview of truncation prescriptions}\label{sec:trunc2}

In this section we provide a brief overview of the results concerning tidal disc truncated from binary companions. The semi-major axis of the cavity (or radius if the cavity is circular) $a_{\rm cav}$ can be predicted analytically, using binary-disc interaction theory, or numerically using hydrodynamical simulations.
These latter simulations show that truncation is generally more efficient for dust than for gas, resulting in larger cavities in dust than the corresponding gas features. This is due to the pressureless/inviscid nature of dust grains\footnote{Even though \citet{longarini2023,lynch2024} discuss the role of dust pressure.} and to the dust drift towards gas pressure maxima, which by definition is located at larger radii than the gas truncation radius.

Despite subtle differences among each other, all truncation mechanisms qualitatively share the same dependence on the parameters: the truncation-radius increases with growing $e$ and $q$ of binary (up to $q\sim 0.1$, above which $a_{\rm cav}$ appears to be almost insensitive to the value of $q$); while it decreases with growing mutual companion-disc inclination $i_{\rm d}$, disc aspect ratio $H/R$, disc viscosity $\nu$.

Such dependences have been thoroughly explored by multiple works that have studied numerically the truncation of dust and gas of circumstellar and circumbinary discs for relatively high binary mass ratios \citep{pichardo2005,pichardo2008,duffell2013,hirsh2020,penzlin2024,dittman2024}, and for planets with lower mass ratios \citep{bryden1999,crida2006,duffell2013,rosotti2016,thun2017,facchini2018,zhang2018,chen2021}.

For planetary mass ratios ($q<0.04$), despite a large number of numerical simulations in the literature studying eccentric/inclined planet disc interaction (e.g. \citealp{goldreich2003,dangelo2006, bitsch2013,ragusa2018,chen2021, baruteau2021,zhu2022,fairbairn2022,tanaka2022,chametla2022,scardoni2023, romanova2024}), only \citet{chen2021} discusses the dependence of truncation (in their case, dust gap width) on planet eccentricity, finding a strong degeneracy between planet mass and its eccentricity, as for higher mass ratios.

Focusing on circular planets, some works (e.g. \citealp{rosotti2016,facchini2018,zhang2018}) found a characteristic relation between the gap width $\Delta=kR_{\rm Hill}$ where $R_{\rm Hill}=a_{\rm bin}(q/3)^{1/3}$ is the companion Hill's radius\footnote{This relation is in qualitative agreement with the scaling of the width of the spatial region where resonance overlapping produces chaotic orbits in the three body problem \citep{wisdom1980}, pointing towards a truncation mechanism that appears to be tightly related to orbital stability considerations.}, and $k=4\,\textrm{--}\,8$ a multiplying factor that depends on the level of coupling between dust and gas and on the degree of evolution of the system (more evolved systems feature wider gaps). In this context, \citet{chen2021} found that when the planet is eccentric, if the size of the planet epicycle exceeds the Hill's radius, the epicycle sets the gap width. This implies that the cavity size scales as $a_{\rm cav}=\max(1+\Delta, 1+be)$, where $b$ is a scale parameter $b\sim k$. Systems where more than one planet is present carve a cavity with edge separated by a distance $\Delta$ from the outer planet (e.g. PDS~70, \citealp{bae2019}).

For stellar binary mass ratios, $q\gtrsim 0.04$, two broad categories of truncation prescriptions can be identified: resonant and non-resonant mechanisms (see Appendix \ref{sec:resononreso} for a thorough discussion). Resonant and non-resonant truncation mechanisms both predict cavity sizes $a_{\rm cav}\sim 2\,\textrm{--}\,4\,a_{\rm bin}$, where $a_{\rm bin}$ is the binary semi-major axis, which are in agreement with typical cavity sizes found in numerical works (e.g. \citealp{miranda2017,hirsh2020, ragusa2020,dittman2024,penzlin2024}).
However, these two truncation mechanisms mainly differ for the fact that the cavity size in resonant mechanisms depends on disc viscosity, while in non-resonant mechanisms it does not depend on the disc properties. In general, resonant theory predicts slightly smaller cavities than non-resonant theory.

Adding to the complexity, numerical studies show that binary/planet-disc interactions can increase the eccentricity of the cavity, which in turn influences its size (e.g. \citealp{dangelo2006,kley2006,pierens2013,miranda2017,thun2017,ragusa2020,munoz2020,pierens2020,siwek2023,toci2024,dittman2024,penzlin2024}). This result is fully captured within non-resonant orbital stability framework, in a few numerical (\citealp{holman1999,petrovich2015,georgakarakos2024}) and analytical \citep{shevchenko2015} studies: in particular, the innermost stable orbit surrounding binaries has a semi-major axis that grows with the test particle eccentricity, thus implying larger cavity sizes for larger cavity eccentricities.

The evolution of disc eccentricity has a complicated dependence on the binary properties ($e$ and $q$), disc parameters ($\alpha\,\textrm{--}\,\nu$ and $H/R$; \citealt{dorazio2021,siwek2023,penzlin2024}), self-gravity (e.g. \citealp{franchini2021} finds limited evolution of disc eccentricity in self-gravitating circumbinary discs), and treatment of disc thermodynamics (e.g. \citealp{sudarshan2022}). This complex dependence is reflected in observations, where circumbinary discs have been observed to host quite eccentric (e.g. HD~142527; \citealt{garg2022}) and also circular cavities (e.g. GG~Tau; \citealt{toci2024}). Thus, to properly assess the size of the cavity, one should also account for the dependence on the disc eccentricity (e.g. \citealp{pierens2013,petrovich2015,ragusa2020}), that can be directly measured observationally through geometric (e.g. \citealp{dong2018}) or kinematic (e.g. \citealp{garg2022,ragusa2024}) considerations.

\section{Resonant and non-resonant truncation} \label{sec:resononreso}

As mentioned in Appendix \ref{sec:trunc2}, binary truncation depends on  concurring physical mechanisms that work together to deplete the cavity region: namely, resonant and non-resonant.

In the first (e.g. \citealp{goldreich1980,artymowicz1994,miranda2015}), the gravitational potential of the binary is decomposed into bar-like potentials revolving with pattern frequencies that are integer multiples or rational fractions of the binary one. The main result from studying the perturbative effects of the binary on the disc dynamics is that each term of the expansion of the potential produces a perturbation at a specific radial location in the disc, different for each term of the potential (resonant radii); at these locations, angular momentum and energy are injected into the disc and are transported away through waves. Viscous effects and shock steepening progressively deposit energy and angular momentum in the disc, resulting in an effective torque \citep{goodman2001,crida2006,cimerman2024}. Although the deposition of angular momentum and energy can occur relatively far from the resonance where the wave was launched, for non-extreme mass ratios ($q> 0.001$) of the binary the deposition occurs relatively close to the resonance. By equating the viscous stresses with the resonant flux of angular momentum in the disc, it is possible to define the truncation radius. For this reason, the resonant criterion always predicts truncation at the location of a resonant radius \citep{miranda2015},  with abrupt jumps when the tidal torque of the dominant resonance exceeds the viscous one.

In the second, the gravitational perturbations produced by the binary potential cause orbital distortions in the disc orbital motion that result in orbital destabilisation and in the depletion of the disc material (e.g. \citealp{papaloizou1977,paczynski1977,pichardo2005,pichardo2008}).
At least three separate mechanisms go under the broad category of ``non-resonant truncation'' mechanisms, each differing in the specific processes responsible for the depletion of the cavity: (i) viscosity dependent tidal torque; (ii) orbital stability;
(iii) orbital intersection.

In (i), the perturbation to the disc, due to the summation of all resonant terms in the expansion of the binary potential, produces a tidal wake whose shape depends on the disc viscosity; the wake breaks the axial symmetry and produces a torque on the binary. Vice versa, the binary exerts the same torque on the disc. Since the viscous torque (i.e. the one attempting to close the gap/cavity) and the tidal torque (opening the gap/cavity) both depend on the disc viscosity, the two contrasting open/close torques scale in the same way, resulting in a truncation criterion that is independent of disc viscosity. Although originally predicted as a mechanism for circumstellar disc truncation by \citet{papaloizou1977}, the same approach has been used by \citet{artymowicz1994} to calculate the truncation radius of circumbinary discs surrounding circular binaries ($e_{\rm bin}=0$), for which resonant truncation appears to underestimate the cavity size.

In (ii), the perturbative terms in the expansion of the binary gravitational potential produce regions in the binary surroundings where no stable orbits exist (resonance overlap), resulting in the formation of a cavity around the binary. The orbital stability of test particles in a binary potential have been widely studied in the context of the restricted three-body problem from celestial mechanics, identifying regions in space where no stable orbits are allowed surrounding the binary where P-type or S-type circumbinary planets can be found (e.g. \citealp{holman1999,quarles2018,adelbert2023,georgakarakos2024}).

Concerning (iii), we first note that orbital stability itself is not sufficient to ensure fluid orbits are not depleted. Indeed, in the proximity of the binary there are regions where stable orbits are possible but intersect other stable outer orbits. When dealing with continuous fluid elements, strong shocks are expected to form in regions with intersecting orbits, depleting the material in the area \citep{paczynski1977,rudak1981,pichardo2005,pichardo2008}.

\section{Comparison between \citet{pichardo2005} and the truncation prescription we adopted}\label{appendix:comparison}

\begin{figure}
   \centering
   \includegraphics[width=\columnwidth]{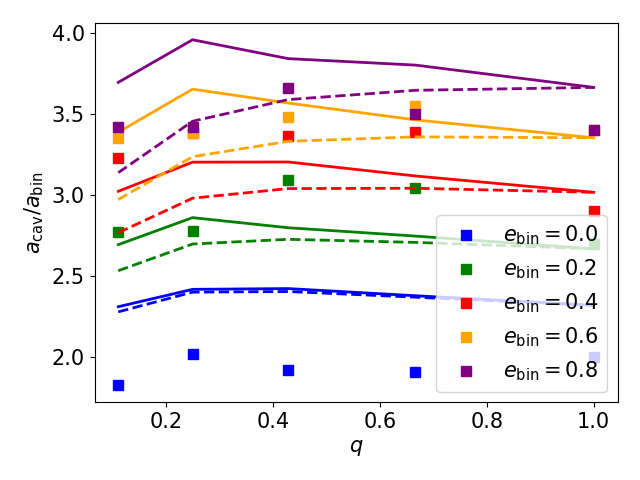}
   \caption{Comparison between the cavity truncation radii reported in \citet{pichardo2005} (data points), and the empirical formula by \citet{georgakarakos2024} in Eq. (\ref{eq:acav}) (solid and dashed lines), for different values of $q$ and $e_{\rm bin}$. Dashed lines assume $e_{\rm cav}=0$, while solid lines use the information about minimum and maximum cavity radii ($R_{\rm min}, R_{\rm max}$) provided by \citet{pichardo2005} to attribute a value of the cavity eccentricity $e_{\rm cav}=(1-R_{\rm min}/R_{\rm max})/(1+R_{\rm min}/R_{\rm max})$, consistent with the percentage shift of the disc centre from the centre of mass reported in \citet{pichardo2008}. The discrepancy between solid lines and squares results in a a reduced chi squared $\tilde \chi^2\sim 1$, assuming a $12\%$ error bar on all data points, and the worst discrepancy is $\sim 15\,\textrm{--}\,20\%$ for the case $e_{\rm bin}=0$.}
\label{fig:pichardotest}%
\end{figure}

Even though the truncation prescription in Eq. (\ref{eq:acav}) has been developed for studying the orbital stability of circumbinary planets and not for disc truncation; we believe it has numerous interesting features that make it particularly suitable for our goal:
\begin{enumerate}[(i)]
\item It reproduces reasonably well the dependence of disc truncation on the binary orbital parameters. This is shown in Fig. \ref{fig:pichardotest}, where we compare Eq. (\ref{eq:acav}) with the disc truncation results of \citet{pichardo2005}, with a good level of agreement between the two. The discrepancy results in a reduced chi squared $\tilde \chi^2\sim 1$, assuming a $12\%$ error bar on all data points, and the worst discrepancy is $\sim 20\%$ for the case $e_{\rm bin}=0$.\\

\item It offers a straightforward estimate of the truncation radius. In contrast, the resonant approach (e.g. \citealp{miranda2015}) requires numerical solution of complex differential equations. Moreover, since resonant mechanisms rely on the strength of the resonant torque at specific resonant locations, $a_{\rm cav}$ exhibits a step-like behaviour, with abrupt jumps when the tidal torque exceeds the viscous one.\\

\item It accounts for the dependence of the cavity size on the mutual binary-disc inclination $i_{\rm d}$, and on disc eccentricity $e_{\rm cav}$. Although the first has been successfully included in resonant theory \citep{miranda2015}, at the moment no works have attempted to include the cavity eccentricity in resonant theory.  \\

\item As previously mentioned, results from non-resonant theory tend to predict slightly larger cavities than those predicted with resonant theory, but do not depend on the disc properties.
This implies that Eq. (\ref{eq:acav}) represents a reasonable estimate of the upper limit on the cavity size for each element of the binary sample and will be treated as such when drawing our conclusions.
\end{enumerate}

We note that \citet{pichardo2008} provides a parametric prescription for circumbinary disc truncation obtained interpolating the dataset presented in \citet{pichardo2005}. But, it does not capture correctly the scaling of $a_{\rm cav}$ for values of $0.01<q<0.1$, which are outside of the $q$ range simulated in those works. Moreover, the prescription in \citet{pichardo2008} does not account for the dependence on binary-disc mutual inclination $i_{\rm d}$, which is instead the case for the \citet{georgakarakos2024} one.

Comparison of solid lines with dashed lines in Fig. \ref{fig:pichardotest} suggests that accounting for disc eccentricity $e_{\rm cav}$ provides a better agreement between data and prescription than $e_{\rm cav}=0$. This result is relevant because a few transition discs have been found to feature eccentric cavities \citep{dong2018,kuo2022,garg2022,yang2023}.

\end{document}